\documentclass[preprint]{aastex}

\usepackage{graphicx}
\usepackage{natbib}
\usepackage{pdflscape} 
\usepackage{amssymb}
\usepackage{ulem} 
\usepackage{color} 

\newcommand{\be}{\begin{equation}}
\newcommand{\ee}{\end{equation}}
\newcommand{\nn}{\mbox{} \nonumber \\ \mbox{} }
\newcommand{\ba}{\begin{eqnarray}}
\newcommand{\ea}{\end{eqnarray}}

\newcommand\eg{{\it{{e.g.,\ }}}}

\newcommand{\Lf}{{Lorentz factor}}
\newcommand{\Bf}{{magnetic field}}

\newcommand{\BH}{{black hole}}

\begin{document}

\title{Early GRB  afterglows from reverse shocks in  ultra-relativistic  long-lasting winds}

\author{Maxim Lyutikov $^1$, Juan Camilo Jaramillo$^{1,2}$\\
 $^1$ Department of Physics and Astronomy, Purdue University, 
 525 Northwestern Avenue,
West Lafayette, IN
47907-2036\\
$^2$ OAN, Observatorio Astron—mico Nacional, Universidad Nacional de Colombia, Bogota, Colombia }

\begin{abstract}
We develop a  model of early GRB afterglows with the dominant $X$-ray contribution from the reverse shock (RS) propagating  in  highly relativistic (Lorentz factor $\gamma_w \sim 10^6$) magnetized wind of a  long-lasting central engine. The model  reproduces, in a fairly natural way,  the overall trends  and yet  allows for variations in the temporal and spectral evolution of early optical and $X$-ray afterglows.  The high energy and the optical synchrotron emission from the RS particles   occurs in the fast cooling regime;  the resulting  synchrotron  power $L_s$ is a large fraction of the wind luminosity, $L_s \approx L_w/\sqrt{1+\sigma_w}$  ($L_w$ and $\sigma_w$ are wind power and magnetization).  Thus, plateaus - parts of  afterglow light curves that show slowly decreasing spectral power  -  are a natural consequence of the RS emission.  Contribution from the forward shock (FS) is negligible in the $X$-rays, but  in the optical both FS and RS contribute similarly:  the FS optical emission is in the slow cooling regime, producing smooth components, while the RS optical emission is in the fast cooling regime, and thus can both produce optical plateaus  and account for  fast optical variability  correlated with the  $X$-rays, e.g., due to   changes in the wind properties. 
 We discuss how the RS  emission  in the $X$-rays and combined FS and RS emission  in the optical can explain many of puzzling properties of early GRB afterglows.
 \end{abstract}
\maketitle 

\section{Introduction}

Gamma Ray Bursts  (GRBs) are produced in relativistic explosions \citep{Paczynski86,piran_04}, that generate a two shock structure - a forward blast wave and the termination shock in the prompt ejecta.
 The standard fireball model \citep{MeszarosRees92,Sari95,PiranReview,MeszarosReview,2015PhR...561....1K} postulates that  (i) the prompt emission is produced by internal collision of matter-dominated ``shells" within the ejecta; (ii) afterglows are generated in the relativistic blast wave after the ejecta deposited much of its bulk energy into circumburst  medium.

One of the most surprising results of {\it Swift}  observations of early afterglows, at times $\leq$ 1 day, is the  presence of unexpected  features - flares and plateaus 
\citep[\eg][]{2006ApJ...642..389N,2013FrPhy...8..661G,2016ApJ...829....7L}.  This challenge the   standard fireball model, that the early $X$-ray are produced in the forward shock \citep{lyutikov_09,2010ApJ...720.1513K}.  

Though the overall hydrodynamic structure of the outflow consisting of the  FS, contact discontinuity and (possibly) a reverse shock is hardly questionable, the observational signatures of shock components are not clear at the moment. In the  {\it  Swift} era, the early afterglow light curves, at times $\leq 10^5$ sec, show numerous variability phenomena, like fast decays, plateaus, flares, various light curve breaks. This variability is  hard to explain within the FS model, which is expected to produce  smooth light curves. Also, the conventional RS signature in matter-dominated models, a  bright flash followed by  a smooth power law decay in the optical band, is rarely observed. 

 Flares and plateaus in 
early $X$-ray   afterglows are interpreted in term of long-lived  central source, which  could remain  active (keeps producing relativistic wind) for long times  after the explosion, $ \sim 10^3-10^4$ seconds, and perhaps even longer \citep[\eg][]{2007ApJ...665..599T}, see \cite{lyutikov_09} for a critique. (We use the term ``afterglow'' in a relation to late, post-prompt, emission, not necessarily coming from the FS.)

In this paper we consider a possibility that most of the $X$-ray afterglow  emission comes from the RS of a long-living central engine, not the FS. The FS shock is there, but it might not radiate.
Previously,  \cite{1998PhRvL..80.1580K,2000ApJ...532..286K,Uhm,2007MNRAS.381..732G,2007ApJ...658L..75G} advocated that  $X$-ray afterglows may come from the reverse shock  (RS) emission. \citep[][argued that radio and millimeter afterglows of GRB 130427A  come from the RS.]{2013ApJ...776..119L} All these works assumed a ``shell paradigm'' for the ejecta - a collection of colliding  (presumably matter-dominated) blobs with mild {\Lf}s. In contrast, in  this work we consider long-lived central engine that produces very fast wind, with the {\Lf}s much larger than that of the forward shock (FS). Building on the  previous ideas outlined in 
\cite{lyutikov_09} we consider temporal and spectral evolution of the RS emission 
and  discussed how various problems with early $X$-ray afterglows can be resolved with the reverse shock paradigm.

\section{Extended source activity: highly relativistic, highly magnetized wind}

The primary GRB explosion may be driven by highly magnetized central source \citep{2006NJPh....8..119L}.
The long-lasting wind from the central source  is even more likely to be magnetized and  highly relativistic, with {\Lf}s much larger than the  \Lf\  of the primary shock. 
In the present work we do not specify the nature of the long-lasting central engine. All three types of the proposed central engines may do: (i) magnetar  \cite{Usov92,2011MNRAS.413.2031M}; (ii) \BH\ jet powered by late accretion \citep{2009MNRAS.397.1153K,2010MNRAS.401.1644B,2011MNRAS.417.2161B}; (iii) BH keeping its magnetic field for long period of time \cite{2011PhRvD..84h4019L,2013ApJ...768...63L}.  In all these cases it is expected that late activity produces fast, clean, magnetized outflows \citep[the primary explosion can be highly magnetized as well][]{2006NJPh....8..119L}. Connecting the present results to a particular model of the central engine, and, thus, to the problem of prompt emission is beyond  the scope of the this work.

The interaction of the wind with the shocked circumburst medium creates a second forward shock (2ndFS below) in the already shocked blast wave,
 a contact discontinuity (CD) separating the wind and the shocked circumburst medium, and the reverse shock in the ejecta.
The wind pushes the contact  discontinuity with time towards the primary FS.  After the external medium absorbs an amount of energy from the wind comparable to the initial explosion, the overall dynamics follows a self-similar blast wave with energy injection \citep[][B\&Mc afterwards]{blandford_76}. But before this time 
 the dynamics of the RS and the primary  FS are very different: the primary FS decelerates as it sweeps more material, while the RS, CD and the second FS propagate though a very tenuous tail of the blast wave, pushed by the wind.  As a result, the {\it  \Lf\ of the RS/CD/2ndFS  combination decreases slower}, see \S \ref{dyn}. This naturally leads to flat emission profiles - the plateaus.

At the same time, the wind produced by the central source can be highly relativistic, so that the RS is always relativistically strong and accelerates particles to high {\Lf}s, of the order of the relative \Lf\ between the RS and the wind. In addition, the wind
can be highly  magnetized, leading to efficient production of synchrotron radiation. All these factors: high bulk \Lf, high \Lf\ of the accelerated particles (in the flow frame) and high magnetization are very beneficial for the production of high energy radiation. Also, in the fast cooling regime a large fraction of the wind luminosity is radiated, while  short cooling times are encouraging to explain the short duration flares.

Finally, we know that reverse  shocks in relativistic outflows do accelerate particles. \cite{2016MNRAS.456..286L,2015MNRAS.454.2754Y} clearly demonstrated that the Inner Knot of the Crab Nebula is the surface of the relativistic termination shock.

\section{Point explosion followed by fast  wind: approximately self-similar dynamics at early times}
\label{dyn}

\subsection{Reverse shock in fast wind}

\begin{figure}[h!]
 \centering
 \includegraphics[width=.99\columnwidth]{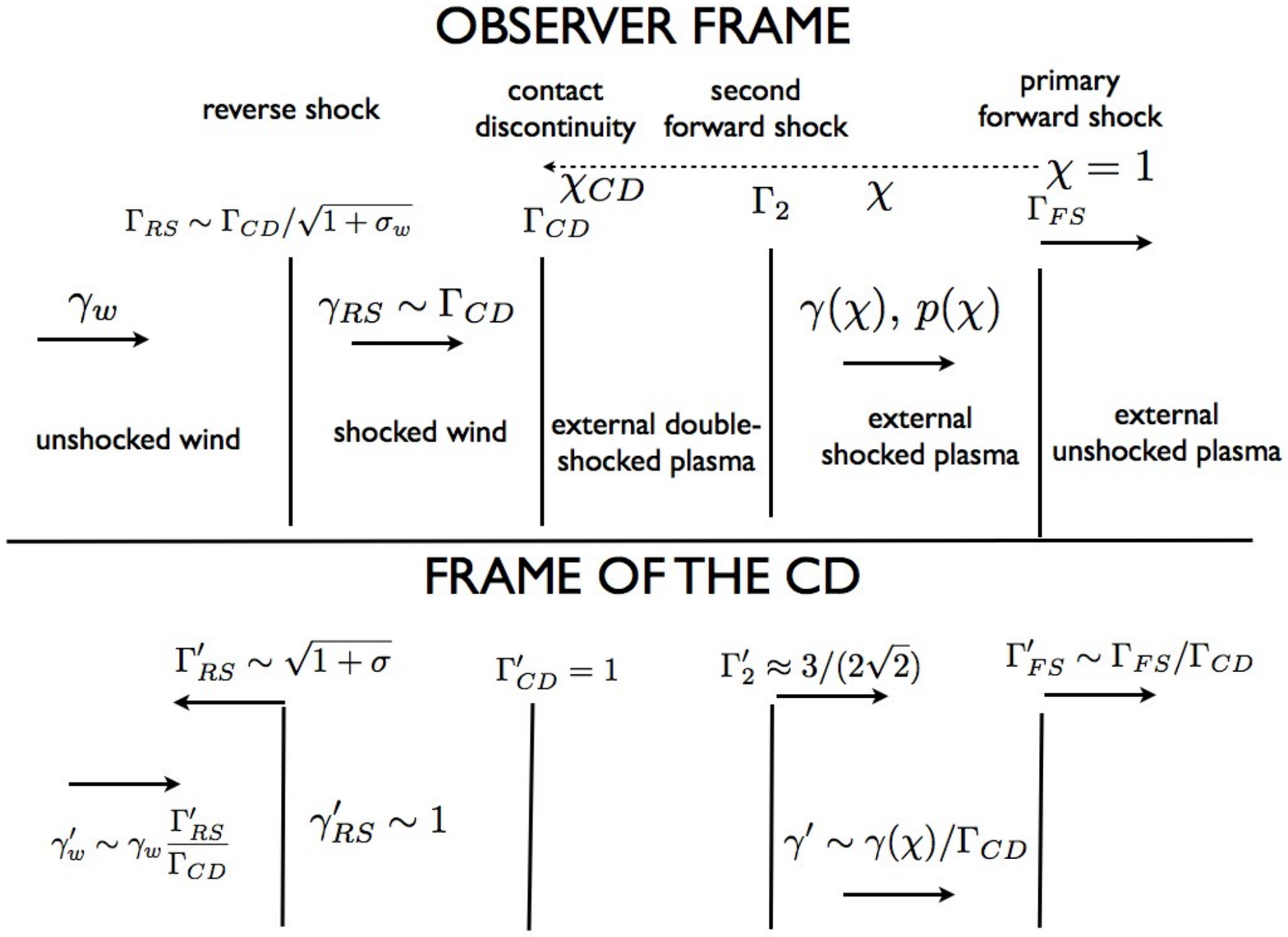}
  \caption{Cartoon of the problem. Top panel: view in the observer frame, lower panel: view in the frame of the CD. The initial explosion  generated a  primary shock propagating  with \Lf\ $\Gamma_{FS}(t)$. The post-primary shock flow ($\gamma,\, p $) depends on the  self-similar coordinate $\chi$ of B\&Mc.  The wind with \Lf\ $\gamma_w$ launches a second shock in the external medium; at the same time  a reverse shock is formed in the wind. The flows are   separated by the contact discontinuity. In the frame of the RS the wind's \Lf\ is $\gamma_w' \sim \gamma_w/\Gamma_{RS}\approx \gamma_w \Gamma_{RS}'/\Gamma_{CD}$.
}
 \label{Shock-structure-RS}
\end{figure}

The initial  explosion with isotropic equivalent energy $E_1$ creates an ultra-relativistic blast wave propagating with $\Gamma_{FS}$, Fig. \ref{Shock-structure-RS}.
\be
\Gamma_{FS} = \frac{1}{2} \sqrt{\frac{17}{2 \pi }}\sqrt{\frac{E_1}{c^5 \rho }}t^{-3/2}
= \left(\frac{17 E_1}{8 \pi  c^5 \rho  t_{{ob}}^3} \right)^{1/8}
\ee
where $ t_{{ob}} = t/\Gamma_{FS} (t)^2$ is the observer time associated with the FS.

The post-primary shock structure is self-similar (B\&Mc).
Suppose that the initial explosion  is followed by a wind with  constant luminosity $L_w$ (generalization to other temporal behavior of $L_w$ are straightforward, but, perhaps are not interesting: we do not expect that the wind power increases with time, while the decreasing wind luminosity will make the effects considered less important).  The wind launches a second forward  shock in the medium already shocked by the primary blast, and the reverse shock in the wind; they are separated by the contact discontinuity (CD).
We expect the dynamics of this triple-shock structure (triple: primary forward shock, second forward shock and the reverse shock, Fig. \ref{Shock-structure-RS}) to have two stages.
At the  early stage the energy deposited by the wind into the shocked media is much smaller than the initial explosion. In this case the CD is located far downstream the first shock; moving  with time in the self-similar coordinate $\chi$ towards the first shock. The motion of the first shock is unaffected by the wind at this stage. After the wind has deposited energy comparable to the initial explosion at time  $t_{eq}$ the whole flow approaches Blandford-McKee self-similar solution with  energy supply.

We assume that the  wind is  magnetized  with luminosity
\begin{equation}
L_{w}=4\pi\gamma_{w}^{2}\left(n_{w}m_{e}c^{2}+\frac{b_{w}^{2}}{4\pi}\right)r^{2}c
\end{equation}
where $n_{w}$ and $b_{w}$ are density and \Bf\ measured  in the wind rest frame. Thus
\begin{equation}
b_{w}=\sqrt[]{\frac{\sigma_w}{1+\sigma_w}} \ \sqrt[]{\frac{L_{w}}{c}}\frac{1}{r\gamma_{w}}
\label{lbw}
\end{equation}
where
\begin{equation}
\sigma_w=\frac{b^{2}_{w}}{4\pi n_{w}m_{e}c^{2}}.
\end{equation}
is the wind magnetization parameter.
We assume that the wind is very fast,  $\gamma_ w \gg \Gamma_{FS},\, \Gamma_{CD}$.

 Interaction of the wind with the CD produces the reverse shock.  The post-RS wind has 
\be 
\gamma_{RS} \approx  \Gamma_{CD}
\ee
(We use capital letters to denote the {\Lf}s of the special surfaces, \eg FS, CD and the RS, while lower cases to denote the flow velocity; primes denote quantities measured in the post-RS flow frame.)

The RS propagates with respect to the CD with velocity $\beta_\sigma$ given by the  post-shock velocity of magnetized relativistic shock \citep[][ Eq. (4.11)]{kennelCoroniti84}.
\be
\beta_\sigma=\sqrt{1-\frac{16 (\sigma_w +1)}{8 \sigma_w ^2+\sqrt{(2 \sigma_w +1)^2 \left(16 \sigma_w ^2+16 \sigma_w +1\right)}+26 \sigma_w +17}}
\ee
 In the observer frame
\begin{equation}
\Gamma_{RS}=\Gamma_{CD}\ \sqrt[]{\frac{1-\beta_{\sigma_w}}{1+\beta_{\sigma_w}}}.
\end{equation}
Neglecting  numerical factors of the order of unity we can approximate
\be
 \Gamma_{RS} \approx \frac{\Gamma_{CD}}{ \sqrt{2} \sqrt{1+\sigma_w}}
 \ee
(for $\sigma_w=0$ and fully self-similar structure this is the exact result; a more correct approximation would be $\sqrt{1+ \sigma_w} \rightarrow  \sqrt{1+2 \sigma_w}$ - we neglect this small difference for the sake of simplicity). 

In physical coordinate $r$, the FS and the RS remain very close,  $r_{RS}\approx r_{FS}\approx  c t$, 
\begin{equation}
\frac{\Delta r}{r_{FS}}=\frac{r_{FS}-r_{RS}}{r_{FS}}
\approx \frac{1+\sigma_w}{\Gamma_{FS}^{2}}
\end{equation}
Only for $\sigma_w \sim \Gamma^2$ (sub-Alfvenic wind) this becomes of the order of unity.

For sufficiently strong wind, with time, the second FS approaches the primary FS. We can estimate the catch-up time by noticing that the power deposited by the wind in the shocked medium scales as $L_w/\Gamma_{CD}^2$. 
 Thus,  
in coordinate time the wind deposits energy similar to the initial explosion at time at time when $\Gamma_{CD} \sim \Gamma_{FS}$,
\be
t_{eq}= \Gamma_{FS}^2 \frac{E_1}{L_w} \approx \left(\frac{E_1^2}{c^5 \rho  L_w} \right)^{1/4}
\ee
This corresponds to the observer time
\be
t_{eq, ob} = t_{eq}/ \Gamma_{FS} ^2(t_eq) = \frac{E_1}{L_w}
\label{teq}
\ee
After time (\ref{teq}) the two forward shocks merge, see \S \ref {later}.
 
\subsection{Self-similar structure of second shocks}

The  fast wind propagating with $\gamma_w$ will launch a second  forward shock  (2ndFS) in the already shocked medium. The wind is separated from the double-shocked external plasma by a contact discontinuity (CD); there is also the reverse shock in the wind. With time the system of RS/CD/2ndFS
may catch up with the first FS.
 The motion of the CD is {\it approximately} self-similar  at early time, well before the catch-up time  (\ref{teq}), and well after.
There are several self-similar regimes for the motion of the  RS/CD/2ndFS, depending on the initial set-up (\eg importance of time delay) and the dynamical approximations, see Appendices \ref{Comp} and \ref{delayed}. Though the details of dynamics are somewhat different in each case they all follow similar patterns.

 Let's first assume that there is no time delay between the first explosion and the onset of the wind (for the applicability  of  the no time delay condition see Appendix \ref{delayed}).
The motion of the CD is  expected to be self-similar before it catches with the primary FS. Let's assume it scales as $\Gamma_{CD} \propto t^{-m/2}$; then the second shock/CD are located at self-similar coordinate  associated with the primary shock (B\&Mc)
\be
\chi_{CD}= \frac{\Gamma_{FS}^2}{ \Gamma_{CD}^2}\propto t^{m-3}
\label{rCD}
\ee
In contrast to fully self-similar solution of B\&Mc,  in which case $\chi_{CD}$ is constant, now it depends on time, $\chi_{CD}(t)$. 

To estimate the motion of the CD, lets' assume 
 $\Gamma_{CD} $  is much larger than the \Lf\ of the local post-first shock flow,  $\Gamma_{CD} \gg {\gamma}(\chi_{CD})$ (to be confirmed  later, see Eq. (\ref{ss})). In the thin shell approximation all the material that passed though the second shock will be concentrated near the CD (hence we neglect the difference between the velocities of 2ndFS and the CD, a common draw-back of the thin-shell approximation,   \citep[\eg][]{Bisnovatyi-KoganSilich95}).  Since the second shock propagates through very hot post-first shock plasma, the energy density in front of the second shock is dominated by the  enthalpy. The total energy in the  post-second shock flow is the total accumulated enthalpy $w_2$,
 \be
  w_2 = 16 \pi \int r^2 p dr= \frac{4\pi}{3} m_p c^5 n_{ex} t^3 \int_{\chi_{CD}}^\infty \frac{d\chi}{\chi^{17/12}}=
\frac{16 \pi}{5} \frac{ \rho c^5 t^3}{\chi_{CD}^{5/12}}, 
  \label{w2}
  \ee
times $\Gamma_{CD}^2$,
\be
E_2 \approx w_2 \Gamma_{CD}^2
\label{E1}
 \ee
 This energy should equal the momentum flux through the RS in the wind, integrated over time
 \be
 E_2 = \int \frac{L_w}{ 4 \pi r^2 c \Gamma_{CD}^2} 4 \pi r^2  c dt = \frac{L_w t}{(m+1) \Gamma_{CD}^2}
 \label{E2}
 \ee
 Thus, 
 \be
 \Gamma_{CD} = \frac{1}{2} \left(  \frac{ 5 }{\pi (m+1)}\right)^{1/4} \frac{L_w^{1/4} \chi_{CD}^{5/48}}{c^{5/4} \rho^{1/4} \sqrt{t}}
\label{Gamma1}
\ee

Solving (\ref{rCD})  and (\ref{Gamma1})
for $\Gamma_{CD}$ and  $\chi_{CD}$ , we find
\ba &&
\Gamma_{CD}=0.50 \frac{E_1^{5/58} L_w^{6/29}}{c^{85/58} \rho ^{17/58} t^{39/58}}
\nn &&
\chi_{CD}= 2.68 \left(\frac{E_1}{c^{5/2} \sqrt{\rho } t^2 \sqrt{L_w}}\right){}^{24/29}
\ea
thus, $m= 39/29$.
The observer time for the CD is
\be
t_{ob} = \frac{t}{2 (m+1) \Gamma_{CD}^2}= \frac{29 t}{136 \Gamma_{CD}^2}
\ee
Thus,
\be
t= 1.08 \frac{E_1^{5/68} L_w^{3/17} t_{{ob}}^{29/68}}{c^{5/4} {\rho }^{1/4}}
\ee
In terms of the observer time,
\ba &&
\Gamma_{CD}=0.48  \frac{E_1^{5/136} L_w^{3/34}}{c^{5/8} {\rho }^{1/8} t_{{ob}}^{39/136}}
\nn &&
\chi_{CD}=\left(\frac{17}{5}\right)^{12/17} \left(\frac{E_1}{L_w t_{{ob}}}\right){}^{12/17}
\label{GammaCD}
\ea
 With time $\chi_{CD}$ decreases -  the CD is getting closer to the FS.
When $t_{ob} \sim t_{eq} = E_1/L_w$ the second shock catches with the primary shock.

\begin{figure}[h!]
 \centering
\includegraphics[width=.99\columnwidth]{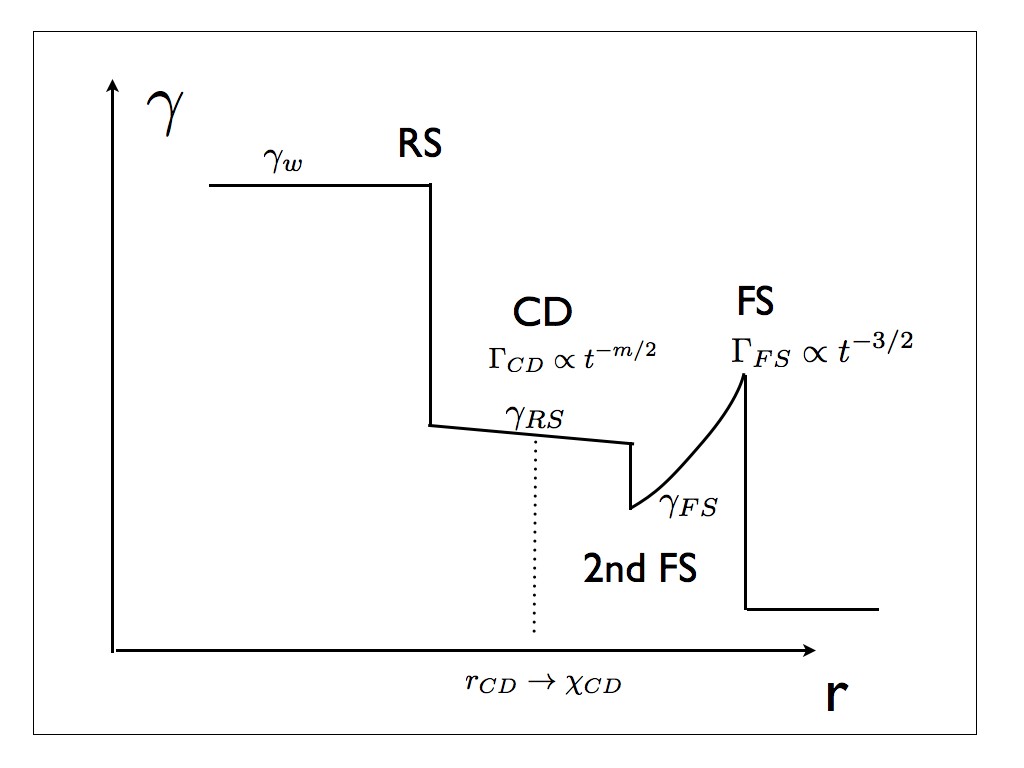}
  \caption{Velocity structure of the triple-shock configuration.  Leading is the FS, that generates a  self-similar post-shock velocity and pressure profiles. The fast wind with \Lf\ $\gamma_w$ is terminated at the RS; the post-RS flow connects through the CD (dotted line)  to the second shock driven in the already shock media (this part of the flow is not considered in detail and is assumed to constant). The CD is located at $r_{CD}$, corresponding to $\chi_{CD}$. The \Lf\ of the flow between the RS and the 2nd FS is expected to slowly decrease with radius; in the present work is it approximated as a  constant \Lf, so that $\gamma_{RS} \approx \Gamma_{CD}$. The RS and 2nd FS are located close to $\chi_{CD}$.
}
 \label{Shock-structure-gamma}
\end{figure}

The \Lf\ of the CD decreases much slower than that of the FS:  $\Gamma_{CD} \propto t^{-39/58} $ compared to $\Gamma_{FS} \propto t^{-3/2}$. (The difference in observer time is not that large, though, $\Gamma_{CD}\propto t_{{ob}}^{-39/136} $ vs $\Gamma_{FS} \propto t_{ob}^{-3/8}$).

We can also verify that the  \Lf\ of the CD (\ref{GammaCD}) is larger than the local \Lf\ of the post-primary shock flow in front of the CD. For $\gamma_{CD} \sim \Gamma_{FS}/(2 \chi_{CD})$ we find
\be
\frac{\Gamma_{CD}}{\gamma_{CD}}=  \left( \frac{17}{5} \frac{E_1}{L_w t_{ob}} \right)^{6/17} \gg 1 \mbox{ for }  t_{ob} \ll  t_{eq}
\label{ss}
\ee
Thus, the wind will drive a second shock into already shocked external medium, see Fig. \ref{Shock-structure-gamma}.

This approach, to treat the second shock/CD in a self-similar way,  has limitations.
 We should  require then that the CD advances with time to smaller $\chi$. For a source luminosity scaling as $L \propto t_{em}^q$, we find that this requires $q>-1$.

\section{Synchrotron emission from self-similar RS}

\subsection{Estimate of efficiency}

Let us first point out the  most important result: high efficiency of the RS emission in the fast wind. Following the treatment of the RS emission in pulsar winds by  \cite{kennel_84} we assume that the RS accelerate all the incoming particles to typical \Lf\ $\gamma_e' \sim \gamma_w'$ - the \Lf\ of the wind in the RS frame (plus an energetically subdominant power-law tail  $f(\gamma') \propto \gamma^{\prime -p}$; the standard choice of $p$ is $2.5$).
The total  synchrotron power $L_s$  is the number of emitting leptons times the power of each electron. 
\be
L_s \approx N P_s
\ee
In the  fast cooling regime the number of emitting electrons is the injection rate $\dot{N}$ times the cooling time $\tau_c$. The cooling time is the energy of an electron divided by the synchrotron power, $\tau_c = \gamma_e m_e c^2/P_s$, where $\gamma_e \approx \gamma_e' \Gamma_{CD}$ is the electron \Lf\ in the observer frame.
Thus 
\be
L_s \approx \dot{N} \tau_c P_s = \dot{N} \gamma_e m_e c^2
\ee
Estimating the injection rate as 
\be
\dot{N} = \frac{L_w}{(1+\sigma_w) \gamma_w m_e c^2}
\ee
gives
\be
L_s \approx \frac{\gamma_e L_w}{(1+\sigma_w) \gamma_w}
\label{Ls}
\ee
Since the minimum lepton \Lf\ in observer frame is 
\be
\gamma_e = \gamma_{w}^\prime \Gamma_{CD} = \left( \frac{\gamma_w}{2 \Gamma_{RS}} \right)  \Gamma_{CD} = \frac{\gamma_w \sqrt{ 1+\sigma_w}}{\sqrt{2 }}
\ee
The total luminosity is then
\be
L_s = \frac{L_w}{\sqrt{2 (1+\sigma_w)}}
\label{Ls1}
\ee

We arrived at an important results: the RS emission is just the wind power (divided by  $\sqrt{1+\sigma_w}$). This could be considered a  natural result - in  fast cooling regime the luminosity is a fraction of the dissipated wind power. Nearly constant winds will produce nearly constant light curve. These are the plateaus.

A relatively weak suppression of the emission in the limit of high magnetization, $\propto \sigma_w^{-1/2}$, is also important; let us elaborate.
 At the RS,  only the part  of the kinetic energy $  \gamma_w' \sim \gamma_w/\Gamma_{RS}$ is converted into photons in the fast cooling regime. This is measured in the  frame of the post shock flow, which moves with \Lf\ $\approx \Gamma_{CD}$ with respect to the observer frame. Thus, the photon energy is  subsequently boosted by $\sim \Gamma_{CD}$. Since $\Gamma_{RS} \sim  \Gamma_{CD}/(\sqrt{2(1+\sigma_w)}$, the emitted power is the fraction $1/{\sqrt{2 (1+\sigma_w)}}$ of the incoming power, (\ref{Ls1}).

The fairly mild suppression of emissivity in high $\sigma_w$ flow is  due to the fact that  in case of higher $\sigma_w$ the reverse shock propagates faster away from the CD  and thus faster towards the wind,  increasing the \Lf\ of the wind in the RS frame. This leads to higher compression at the shock, higher {\Lf}s of the emitting particles, mostly  off-setting the commonly known problem of low efficiency in high-$\sigma_w$ shocks  \cite[for comparison, in the frame of the shock, only $\sim 1/(8 \sigma_w^2)$ fraction of incoming energy is dissipated,][]{kennelCoroniti84}.

\subsection{Typical frequencies}

Next we estimate the corresponding synchrotron  peak frequency.
The \Lf\ of the wind in the frame of the RS is 
\be
\gamma_w^\prime = \frac{\gamma_w}{2 \Gamma_{RS}}= \frac{\gamma_w \sqrt{1+\sigma_w}} { \sqrt{2}  \Gamma_{CD}}
\label{gammawprime}
\ee
Using (\ref{lbw}) for \Bf\ in the wind, 
in the post-RS region the \Bf\ is 
\be
b_{RS}= \frac{\gamma_w}{ \Gamma_{CD}} b_w=\frac{ \sqrt{L_w}}{\sqrt{c} r \Gamma_{CD}} \sqrt{\frac{\sigma_w }{\sigma_w +1}}
 \ee
Synchrotron photon energy at the break (assuming $\gamma_e' \sim \gamma_w^{\prime} $) is 
\ba &&
\epsilon_m = \hbar \gamma_{RS} \gamma_w^{\prime 2}  \frac{e b_{RS}}{m_e c}=
\frac{e \hbar}{2 c^{5/2} m_e} \frac{L_w^{1/2} \gamma_w^2   \sqrt{\sigma_w  (\sigma_w +1)} }{t \Gamma_{CD}^2}=
\label{epsilons}
\\ &&
2 \sqrt{\frac{17}{145}}  \sqrt{\frac{m_p}{m_e} }  \hbar   \sqrt{\frac{4\pi n_{ex} e^2}{m_e} } \frac{\sqrt{\sigma_w  (\sigma_w +1)} \gamma _w^2}{\chi _{{CD}}^{5/24}}
= 650 \sqrt{\sigma_w  (\sigma_w +1)} 
\frac{L_{w,50}^{5/34} t_{{ob},3}^{5/34} \gamma _{w,6}^2}{E_{1,54}^{5/34}}  \, {\rm eV}
   \label{epsilons1}
   \ea
where  $n_{ex}$ is the external medium number density,  and numerical estimates assume that the quantities are measured in cgs units; we use notation $X_n\equiv X/10^n$. The energy (\ref{epsilons1}) is the minimum energy - the break energy -   acceleration of particles at the RS will produce a cooled power-law emission  above  (\ref{epsilons1}), stretching, presumably, to hundreds of keV.
Note that $\epsilon_m$ increases with time - this is due to the fact that the CD slows down, so that for constant \Lf\ of the wind $\gamma_w$, the
 relative \Lf\ of wind in the frame of the RS $\gamma_w' \propto \gamma_w/\Gamma_{CD}$ increases.

Emitted synchrotron power per electron of the wind is
\be
P_{s} \approx \gamma_{RS}^2 \gamma_w^{\prime , 2} \frac{e^4  b_{RS}^2}{ m_e^2 c^3}=\frac{e^4 \sigma_w  L_w \gamma _w^2}{2 c^6 t^2 \Gamma _{{CD}}^2 m_e^2}
\ee
The 
cooling time for the electrons emitting at $\epsilon_m$ 
\be
\tau_{c,m}\approx \frac{\gamma_w^{\prime} \gamma_{RS} m_e c^2}{P_s} = \frac{m_e^3 c^8}{e^4 } \frac{t^2 \Gamma_{CD}^2}{\gamma_w L_w } \frac{\sqrt{1+\sigma_w}}{\sigma_w}
\ee
We find
\be
\frac{\tau_{c,m}}{t_{ob}} \approx  3 \times 10^{-2} \frac{E_{1,54}^{15/68}}{{n_{ex}}^{3/4} L_{w,50}^{8/17} t_{{ob},3}^{49/68} \gamma
   _{w,6}} 
   \ee
placing $X$-ray emission, and higher energy  $\gamma$ ray emission, well in the fast cooling regime (after $t_{ob} \sim $ few seconds).

\begin{figure}[h!]
 \centering
\includegraphics[width=.99\columnwidth]{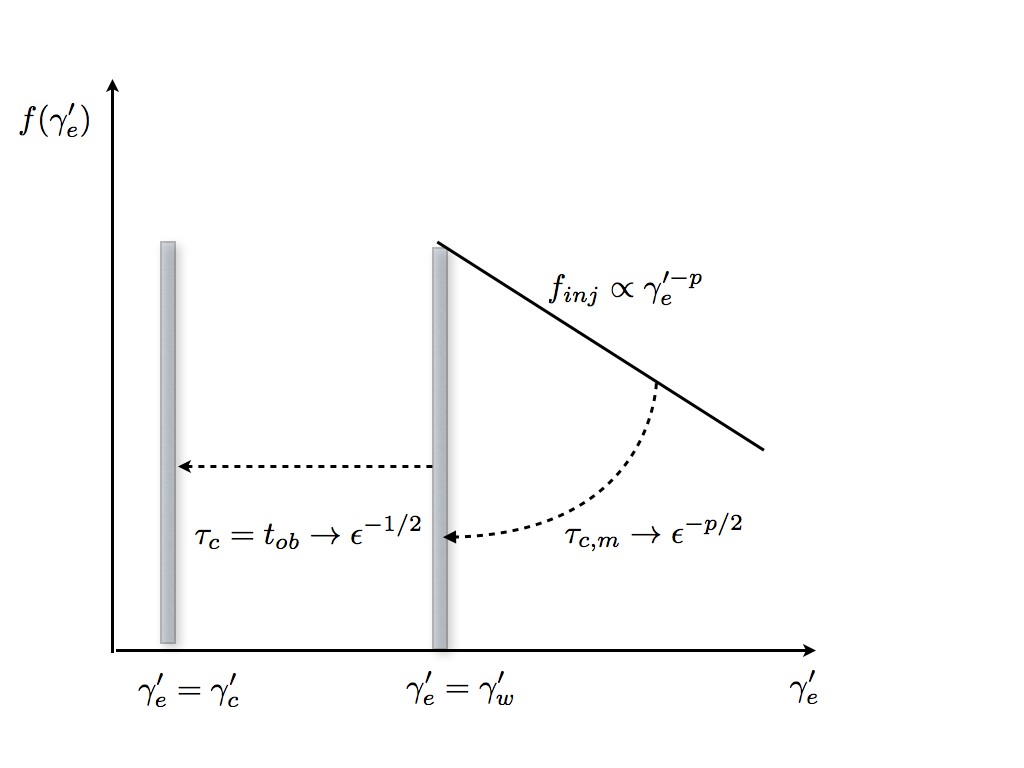}
  \caption{Evolution of the distribution function of the RS-accelerated leptons in the fast cooling regime. The RS injects leptons  with a power law distribution $f_{inj} \propto \gamma_e^{\prime -p}$ and the minimal energy of the order of the wind's \Lf\ in the frame of the RS, $\gamma_e' \sim \gamma_w'$. Quickly, in the observer time  $\tau_{c,m}$, all leptons cool to the lowest energy, producing the cooled spectrum  $\propto \epsilon^{-p/2}$ above $\epsilon_m$; and then, on  observer time  $\tau_{c}\sim t_{ob}$, cool down to the cooling break energy, producing the  spectrum  $\propto \epsilon^{-1/2}$, after \protect\cite{1998ApJ...497L..17S}.}
 \label{Shock-structure-fastCooling}
\end{figure}

At a given observer time $t_{ob}$ we can estimate the photon energy  $\epsilon_{c} $ below which emission would be in a slow cooling region, $\tau_c\approx t_{ob}$.
Regime of fast cooling starts at $\gamma_c$ that satisfies the condition
\ba &&
\gamma_e m_e c^2 \approx \gamma_{RS} \gamma_c m_e c^2= P_{c} t_{ob}
\nn &&
 P_{c} = \gamma_{RS} ^2 \gamma_c^2 \frac{e^4 b_{RS}^2 }{m_e^2 c^3}
 \nn &&
 \gamma_c =0. 1 \frac{m_e c^{39/8}}{e^4} \frac{L_w^{1/2}}{E_1^{7/8} \rho^{5/8}} \frac{1+\sigma_w}{\sigma_w} t_{ob}^{5/8}
 \label{gammac}
\ea
The corresponding energy is
\be
 \epsilon_{c}  = \hbar  \gamma_{RS}^2 \gamma_c^2 \frac{e b_{RS} }{m_e c}\approx  
 0.28 \frac{E_1^{5/17} c^{17/2} m_e^5 (\sigma_w +1)^{3/2} \hbar }{e^7 \rho  \sigma_w ^{3/2}
   L_w^{27/34} t_{{ob}}^{22/17}}=
0.75    
 \frac{E_{1,54}^{5/17}}{{n_{ex}} L_{w,50}^{27/34} t_{{ob},3}^{22/17}}
  \, {\rm eV}
\label{ecc}
\ee
Thus,  optical electrons at the RS emit in the mildly fast cooling regime.
Evolution of the electron distribution function of the RS-accelerated electrons is illustrated in Fig. \ref{Shock-structure-fastCooling}.

\begin{figure}[h!]
 \centering
\includegraphics[width=.99\columnwidth]{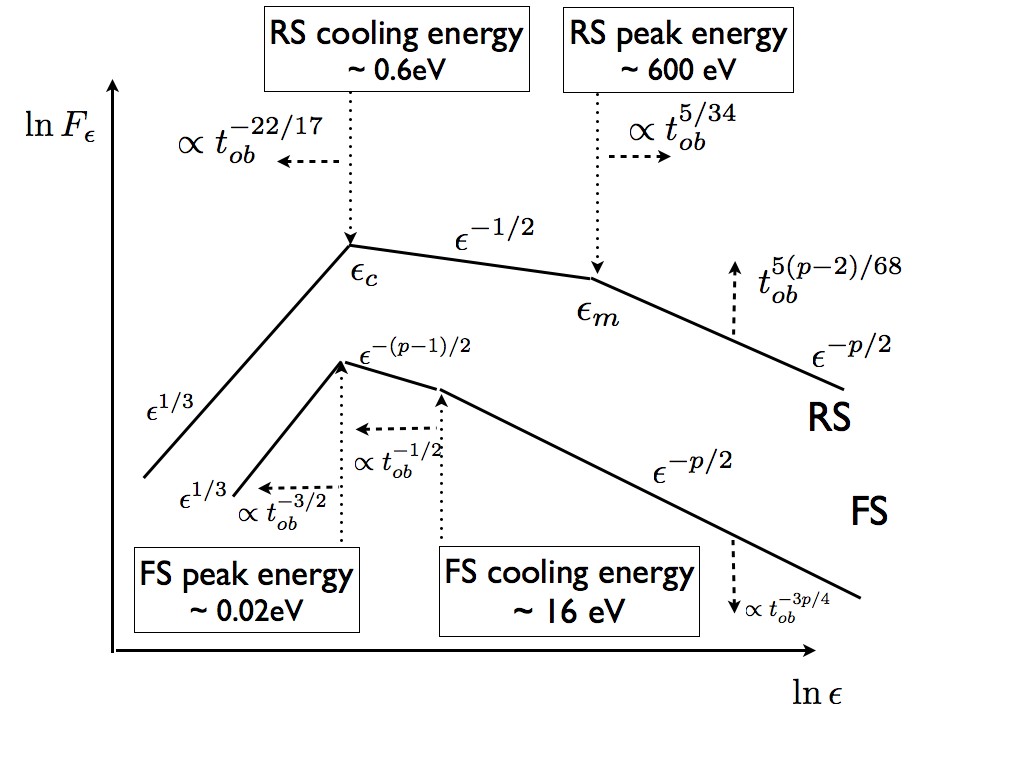}
  \caption{Relative spectra of the reverse shock (RS) and the forward shock (FS) emission estimated at the observer time $t_{ob} =10^3$ seconds. The RS is in a fast cooling regime; the break energy (corresponding to the minimum energy of the accelerated leptons) is $\sim 5$ keV, Eq. (\protect\ref{epsilons}); the cooling energy is  $\sim 10$ eV, Eq. (\protect\ref{ecc}). The FS emission is in the slowly cooling regime -  the cooling energy is  $\sim 0.02$ eV, Eq. (\protect\ref{epsilonsFS}). In the optical, the fluxes are comparable. It is assumed that both the RS and the FS accelerate particles with the same power law index $p$. 
  Dashed arrows indicate temporal evolution of  the break frequencies and fluxes in fixed energy bands (optical and $X$-rays). 
  Numerical values are given for $t_{ob} =10^3 $ seconds.
}
 \label{Shock-structure-spectra}
\end{figure}

In the fast cooling regime the observed flux is 
\ba &&
F_{\epsilon}= \left \{\begin{array}{lcc}
				 F_{\epsilon_m} \ (\epsilon/\epsilon_{c})^{1/3} (\epsilon_c/\epsilon_{m})^{-1/2} ,  & \epsilon<\epsilon_{c} \\
             \\ F_{\epsilon_m}\ (\epsilon/\epsilon_{m})^{-1/2}, & \epsilon_{c}<\epsilon<\epsilon_{m}   \\
             \\ F_{\epsilon_m} (\epsilon/\epsilon_{m})^{-p/2}, & \epsilon> \epsilon_{m}
             \end{array}
             \right.
             \nn &&
   F_{\epsilon_m}= \frac{L_w}{\sqrt{2 (1+\sigma_w)} \epsilon_m} \frac{1}{4 \pi D^2}           
\ea
where $D$ is the distance to the source
(see Fig. \ref{Shock-structure-spectra}).

In this work we concentrate on the optical/high energy emission. We expect that the effects of synchrotron self-absorption are not important.

\section{Spectra and time-evolution}

\subsection{Reverse shock}

Let us calculate the expected spectra and light curves. Below, we restrict our calculations to the case of constant luminosity source and constant external density. The external  wind environment is discussed in Appendix \ref{wind}

In the $X$rays, the peak flux at $\epsilon_m$ is
$ F_{\epsilon,m} \approx {L_w}/{\epsilon_{m}}$. For $\epsilon > \epsilon_{m}$ we have
\be
  F_{\epsilon} = F_{\epsilon,m} \left( \frac{\epsilon}{\epsilon_{m}}\right)^{-p/2} \propto t_{ob} ^{ \frac{5}{68} (p-2)} \approx t_{ob}^{0.036}
  \label{Fnu}
  \ee
  where we assumed $p=2.5$.
  The spectral flux is slowly rising. This is the plateau. 
  
  One of the predictions of the model could be  the correlation between the temporal decay, $  F_{\epsilon}  \propto t^{-\alpha}$, and the spectral index $\beta$,  $  F_{\epsilon}  \propto  \epsilon^{-\beta}$ of plateaus. Since   $\beta = p/2$, 
  \be
   F_{\epsilon}  \propto t_{ob}^{5 (\alpha-1)/34}.
   \ee

In the frequency range between the cooling frequency and the peak frequency, $\epsilon_{c} < \epsilon < \epsilon_m$, the spectrum is $\propto \epsilon^{-1/2}$. The ratio of optical to hard $X$ -rays spectral fluxes from the RS are
\be
\frac{F_{\epsilon ,O}}{F_{\epsilon,X}}= \left( \frac{\epsilon_0}{\epsilon_m}\right)^{-1/2} \left( \frac{\epsilon_m} {\epsilon_X}\right)^{-p/2} \approx 10^3
\label{FOFX}
\ee
for $\left( \frac{\epsilon_0}{\epsilon_m}\right)^{1/2} \sim 30$ and $\left( \frac{\epsilon_m} {\epsilon_X}\right)^{p/2} \sim 30$
($\epsilon_0 \sim 1 $  eV, $\epsilon_m \sim $ few keV, $\epsilon_X \sim $ tens of keV). The total energetics in optical will be a factor $\epsilon_O/\epsilon_X \sim 10^{-4}$ smaller that the estimate (\ref{FOFX}).

The estimate (\ref{FOFX}) compares well with observations. For example, for GRB 130603B \cite[][Fig. 2]{2014ApJ...780..118F},  at  0.47 days the ratio of   optical ($\sim 50 $ $\mu$J) to X-ray  flux ($\sim 5 \times 10^{-2}$ $\mu$J)   is $\sim 10^3$.

\subsection{Forward shock: optical, no $X$-ray}

  The FS models of early afterglows assume that fraction of energy $\epsilon_e$ and  $\epsilon_B$ are converted into accelerated particles and \Bf. The emission from the FS in our model follows the conventional prescription for afterglow emission \citep{1998ApJ...497L..17S}, but with
  smaller  $ \epsilon_B, \,  \epsilon_e \sim 10^{-3}$.  These values, especially much smaller $\epsilon_B$, are better justified, that the usually used ones
of $ \epsilon_B, \,  \epsilon_e \sim 10^{-1}-10^{-2}$,
 \citep[\eg][]{2003ApJ...596L.121S,2008ApJ...673L..39S}.
 The lower values of    $ \epsilon_B, \,  \epsilon_e $  shifts the spectrum down in intensity, the peak to lower frequencies and makes the slow cooled range of energies wider.

The \Bf\ and  the \Lf\ of accelerated particles in the FS region are \citep{1998ApJ...497L..17S}
\ba &&
b_{FS}= \sqrt{32 \pi \rho c^2}  \sqrt{\epsilon_B} \Gamma_{FS}  
\nn &&
\gamma_e = \epsilon_e \frac{m_p}{m_e} \Gamma_{FS}.
\ea
We estimate using our fiducial numbers
\ba &&
\epsilon_{m, FS} = \hbar \Gamma_{FS} \gamma_e^2 \left(\frac{e b_{FS}}{m_e c} \right)=
 \sqrt{ \frac{17}{2} } \frac{ e \hbar m_p^2}{m_e^3 c^{5/2} }\sqrt{\epsilon_B}  \epsilon_e^2 \frac{\sqrt{E_1}}{t_{ob}^{3/2}} = 0.02 \, {\rm eV} \sqrt{\epsilon_{B,-3}}  \epsilon_{e,-3}^2
\frac{E_{1,54}^{1/2}}{t_{ob,3}^{3/2}}
\nn &&
\tau_{c,FS} = \frac{m_e^4 c^3}{32 \pi e^4 m_p} \frac{1}{\Gamma_{FS}^4 \rho \epsilon_e \epsilon_B }
\nn &&
\frac{\tau_{c,FS}}{t_{ob} }=  \frac{c^{11/2} m_e^4 t_{ob}^{1/2} }{4 \sqrt{17 \pi} e^4  m_p \sqrt{ E_1 \rho}  \epsilon_e  \epsilon_B }
=100 \frac{t_{ob,3}^{1/2}}{E_{1,54}^{1/2} n_{ex}^{1/2} \epsilon_{B,-3} \epsilon_{e,-3}}
\label{epsilonsFS}
\ea
Thus the peak emission is in IR, uncooled.

The energy of the particles with ${\tau_c} \sim {t_{ob} }$ is
\ba &&
\gamma_c= \frac{m_e ^3 c^3}{32 \pi e^4 \rho t_{ob} \Gamma_{FS}^3 \epsilon_B}
\nn &&
\frac{\gamma_c}{\gamma_e } = 27 \frac{t_{ob,3}^{1/2}}{E_{1,54}^{1/2} n_{ex}^{1/2}   \epsilon_{e,-3}  \epsilon_{B,-3}^2}
\ea

Emission at the cooling break falls into UV
\be
\epsilon_{c,FS} = \hbar \Gamma_{FS} \gamma_c^2 \left(\frac{e b_{FS}}{m_e c} \right)= 16 {\rm eV}\, {E_{1,54}^{-1/2}} n_{ex}^{-1} \epsilon_{B,-3}^{-3/2} t_{ob}^{-1/2}
\ee
Thus FS produces optical into uncooled regime. Synchrotron emission in the slowly cooling regime is expected to produce  smoother light curves.

The peak spectral power at $\epsilon_m$ remains constant in time  \citep[][Eq. 11]{{1998ApJ...497L..17S}}
\be
F_{\epsilon_m, FS} \approx  ( c t)^3 n_{ex} \frac{P_{m, FS}}{\epsilon_{m,FS} }=\frac{34 \sqrt{2 \pi }}{3} 
\frac{e^3  \sqrt{n_{{ex}}E_1} \sqrt{m_p}}{c^3 \hbar  m_e} \sqrt{\epsilon _B},
\ee
but the spectral flux is decreasing for $\epsilon > \epsilon_m$ at a given observational range 
\be
F_{\epsilon, FS} \propto \epsilon_m^{-p/2} \propto t_{ob}^{-3 p/4} \propto t_{ob}^{-1.875}
\label{FFS}
\ee
(where used the evolution of $\epsilon_{m, FS}$ from (\ref{epsilonsFS}). Sharp decline of the $X$-ray spectral flux in the FS model, Eq. (\ref{FFS}), can be  contrasted with the slow evolution in the RS model, Eq. (\ref{Fnu})

\subsection{Comparing optical and $X$-ray fluxes of FS and RS}

The particles from the high energy tail in the FS   can still emit in $X$-rays. 
The required \Lf\ is 
\be
\gamma_X \approx3 \times 10^5 \frac{t_{ob,3}^{3/8} \epsilon_{X,5}^{1/2}}{E_{1,54}^{1/8} n_{ex}^{1/8}   \epsilon_{e,-3}  \epsilon_{B,-3}^{1/4}} \approx 10^2 \gamma_e
\ee 
The ratio of $X$-fluxes from the RS (\ref{epsilons}) and the FS  (\ref{epsilonsFS}) are (assuming same $p$ for both) 
\be
\frac{F_{X ,FS}}{F_{X,RS}}= \frac{F_{\epsilon_m, FS}}{F_{\epsilon_m}}\left( \frac{\epsilon_{c,FS}}{\epsilon_{m}} \right)^{p/2}
\left( \frac{\epsilon_{c,FS}}{\epsilon_{m,FS}} \right)^{-(p-1)/2}
=4 \times 10^{-3}
\frac{E_{1,54}^{1.16} \epsilon _{B,-3}^{0.125} \epsilon
   _{e,-3}^{1.5}}{{n_{ex}}^{0.125} L_{w,50}^{1.04} t_{{ob},3}^{1.41}}
      \ee
Thus, the $X$-ray contribution from the FS is negligible. 

On the other hand,  the ratio of optical fluxes from the RS (\ref{epsilons}) and the FS  (\ref{epsilonsFS}) is 
\be
\frac{F_{O ,FS}}{F_{O,RS}}=\frac{F_{\epsilon_m, FS}}{F_{\epsilon_m}}\left( \frac{\epsilon_{m,FS}}{\epsilon_{O}} \right)^{(p-1)/2}
\left( \frac{\epsilon_{m}}{\epsilon_{O}} \right)^{-1/2}
=0.16  \frac{E_{1,54}^{1.3} {n_{ex}}^{3/4} \epsilon _{B,-3}^{0.875} \epsilon
   _{e,-3}^{1.5}}{L_{w,50}^{63/68} t_{{ob},3}^{1.05}} \left( \frac{\epsilon_O}{{\rm 1\, eV}}\right)^{-1/4}
\ee
where $\epsilon_O$ is the energy of the optical photons.

We conclude that the FS and the RS can  contribute similarly to optical fluxes (given the uncertainties of the parameters). This is important: it allows us to explain smooth optical light curves (as emission of the FS and constant RS) and possible fast variations at the RS due to changing wind conditions.

\subsection{Light-curves and spectra}

\begin{figure}[h!]
 \centering
\includegraphics[width=.99\columnwidth]{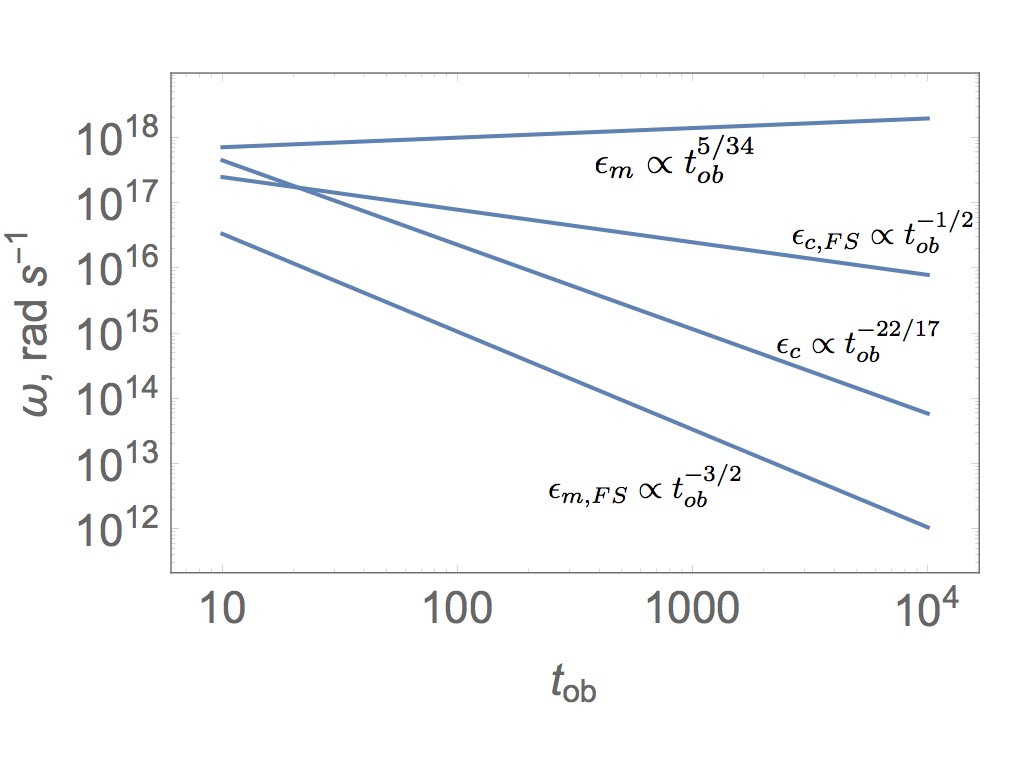}
  \caption{Evolution of typical frequencies (cooling and break) for the RS shock ($\epsilon_m$ and $\epsilon_c$) and the FS   ($\epsilon_{m,FS}$ and $\epsilon_{c,FS}$). (Parameters are $E_1 =10^{54}$ erg, $L_w= 10^{50} $ erg s$^{-1}$, $\gamma_w =10^6$, $\epsilon_e=\epsilon_B= 10^{-3}$, $p=2.5$, $n_{ex}=1$ cc, $\sigma_w =1$; distance to the source is $3 \times 10^9$ pc. These parameters are used in the following figures as well. For the plots we also assume constant luminosity wind and constant external density.)}
 \label{Shock-structure-freq}
\end{figure}

In the present model the RS is highly relativistic, magnetized and produces emission in the fast cooling regime. In contrast the FS is low-magnetized and produces emission in the slow cooling regime. Evolution of typical frequencies  with time for the FS and RS are compared in Fig. \ref{Shock-structure-freq}.

\begin{figure}[h!]
 \centering
\includegraphics[width=.99\columnwidth]{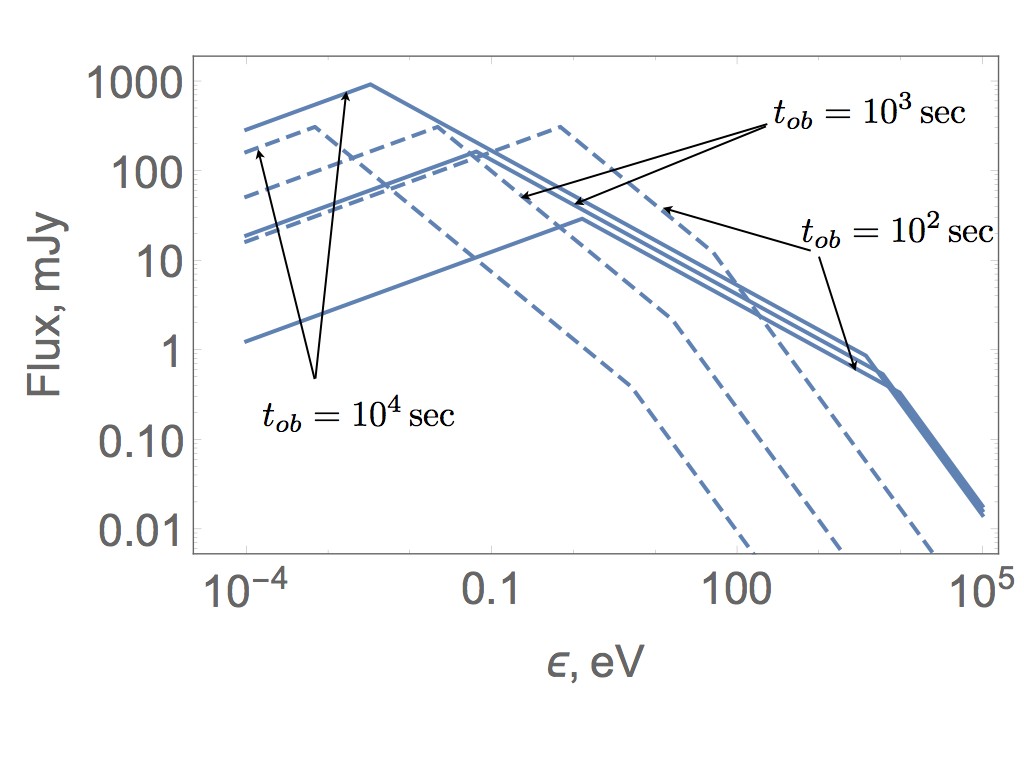}
  \caption{Comparison of spectra of the  RS emission (solid lines)  and  of the FS emission (dashed lines) for $t_{ob} = 10^2, 10^3, 10^4$ sec. In the X-rays the RS emission dominates, while in the optical both contribute approximately equally. With times both sets of curves move to the left, to lower break energies. The fluxes at $X$-rays remain nearly constant - these are the plateaus.}
 \label{compareX0}
\end{figure}

Next, we combine our calculations of the spectra produced by RS and the FS, Fig. \ref{compareX0}. The $X$-ray afterglows are dominated by the RS, while in optical contribution from the FS may also be important.  

\begin{figure}[h!]
 \centering
 \includegraphics[width=.49\columnwidth]{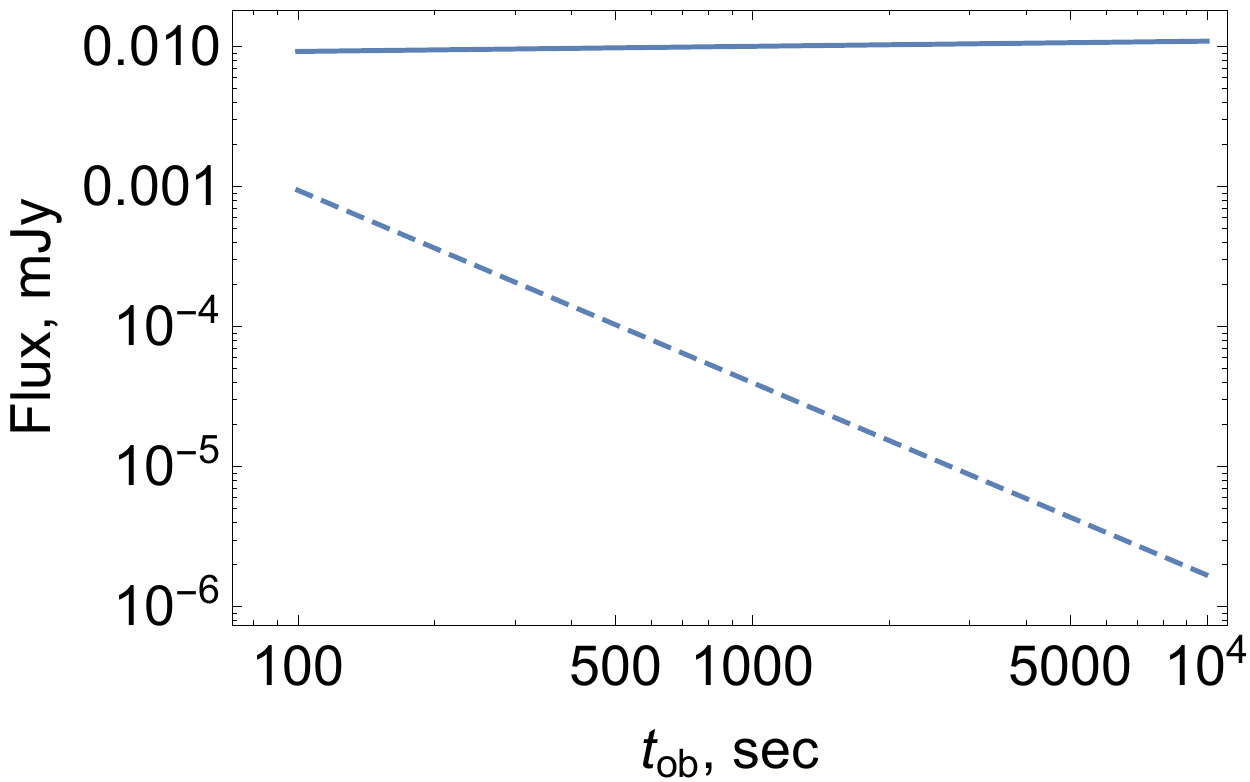}
\includegraphics[width=.49\columnwidth]{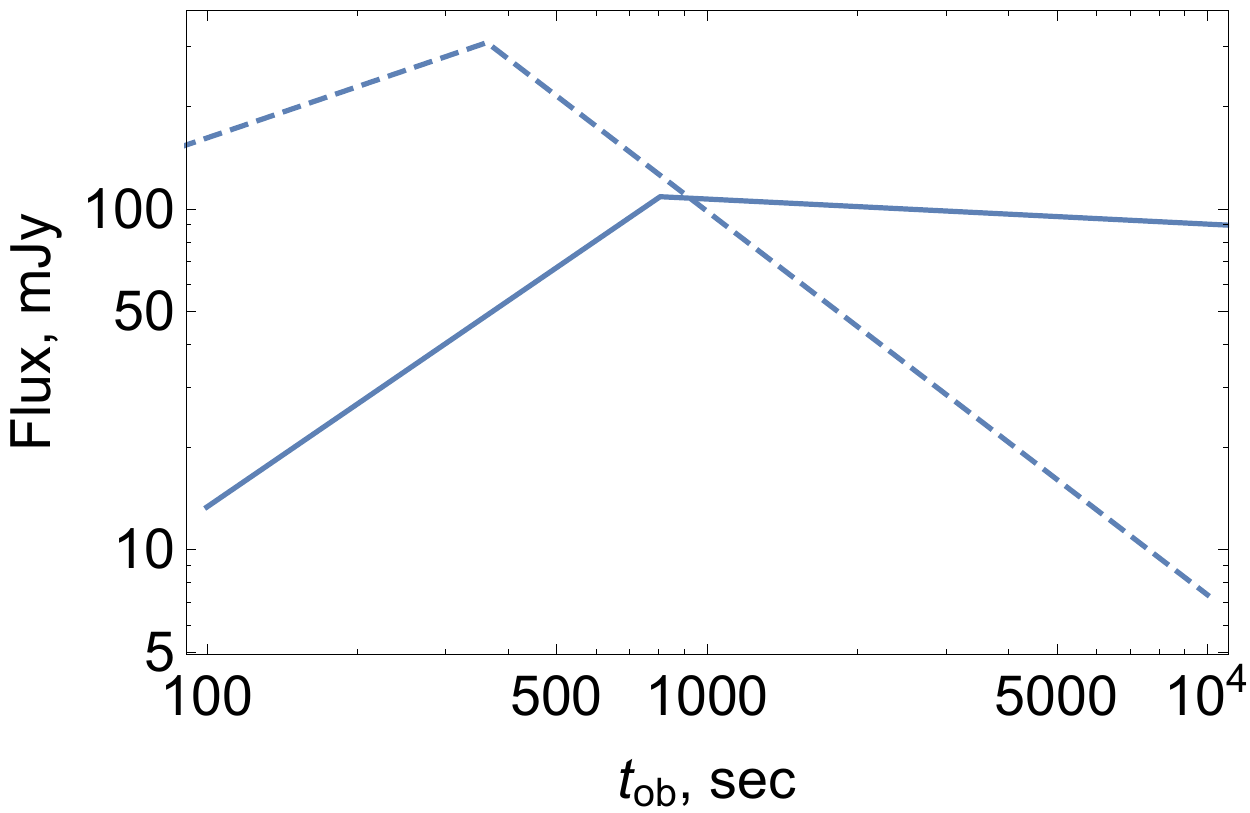}
  \caption{ Comparison of RS emission (solid lines)  and FS emission (dashed lines) as a function of time for $X$-rays ($\epsilon=10^5$ eV, left panel) and optical   ($\epsilon=1$ eV, right panel). In the $X$-rays the RS flux is slowly increasing, Eq. (\protect\ref{Fnu}). At early times the FS may dominate in the optical, but it always subdominant in the $X$-rays. In the optical the break in the RS emission is associated with the cooling break, while in the optical it is  the minimum injection energy. RS also produces plateau in the  optical band. The changing dominance of RS and FS contribution is expected to produce a spectral change from uncooled FS $\propto \epsilon^{-(p-1)/2}$ to cooled RS $\propto \epsilon^{-1/2}$.}
 \label{Fepslion}
\end{figure}

In Fig. \ref{Fepslion} we plot observed fluxes in $X$-rays and in the optical. In the $X$-rays the RS produces very flat light curve, see Eq. (\ref{Fnu}), while the FS emission is fast decaying, Eq. (\ref{FFS}). In the optical both FS and RS emission experience a break near $t_{ob} \sim 100$ seconds. But the nature of the breaks is different: For the RS  it is the passing of the cooling break, while for the FS it is the effect of minimal \Lf. 

To be in fast cooling regime it is required  that $\epsilon_{c} < \epsilon_{m}$. This puts a lower limit on $\gamma_w$:
\be
\gamma_w \geq 0.4 \frac{c^{17/4} m_e ^3}{e^4}
 \frac{E_1^{15/68}}{L_w^{8/17} \rho^{3/4} t_{ob}^{49/68}} \frac{\sqrt{1+\sigma_w}}{\sigma_w} \approx \gamma_w \geq 3 \times 10^4
 \label{gammawmin}
 \ee

\subsection{Variable wind properties}

All the calculations done above are for constant \Lf\  and magnetization of the wind. Variations in these quantities are expected to produce variations in  the  light curves. 
These variation can be taken into account, approximately, if we notice that the time when a particular element of  wind left the central source is approximately the emission observer time.  Thus,  changing wind properties can be approximately taken into account by introducing $L_w(t_{ob})$, $\sigma_w(t_{ob})$,  $\gamma_w(t_{ob})$. 

\begin{figure}[h!]
 \centering
\includegraphics[width=.49\columnwidth]{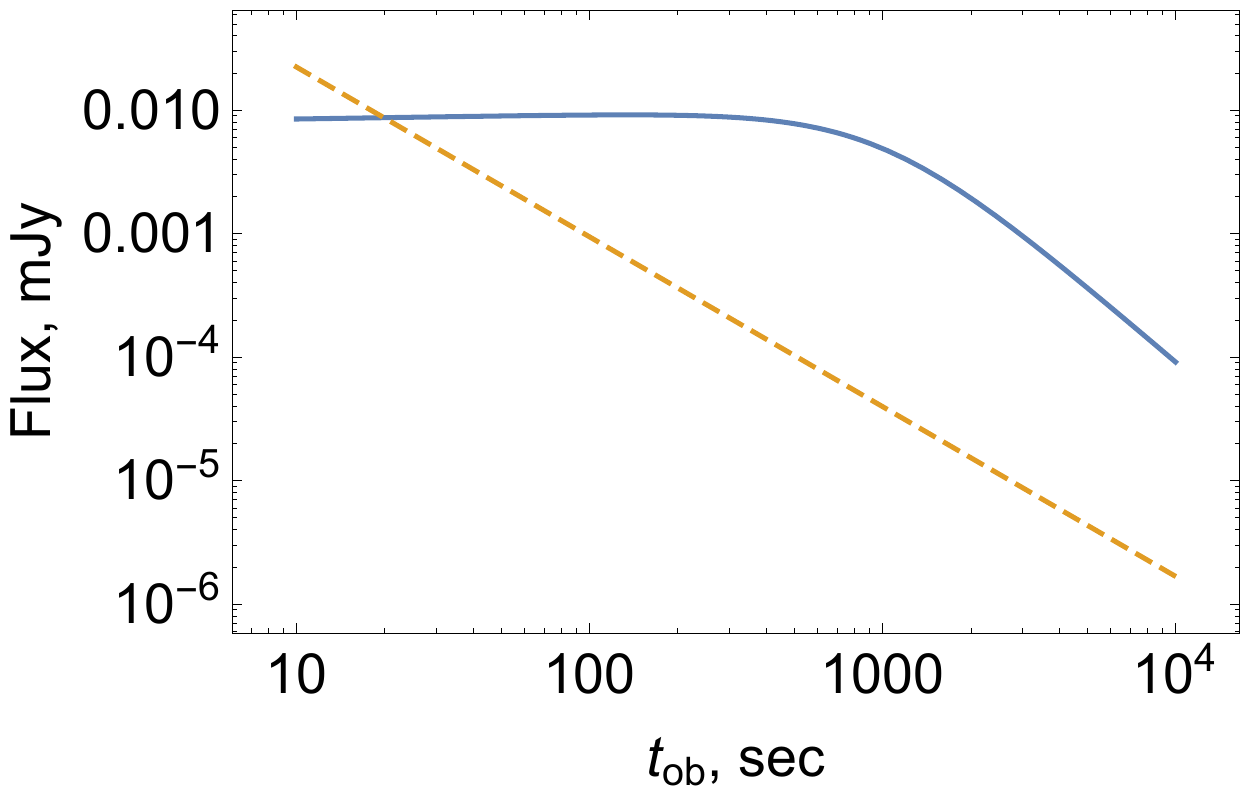}
\includegraphics[width=.49\columnwidth]{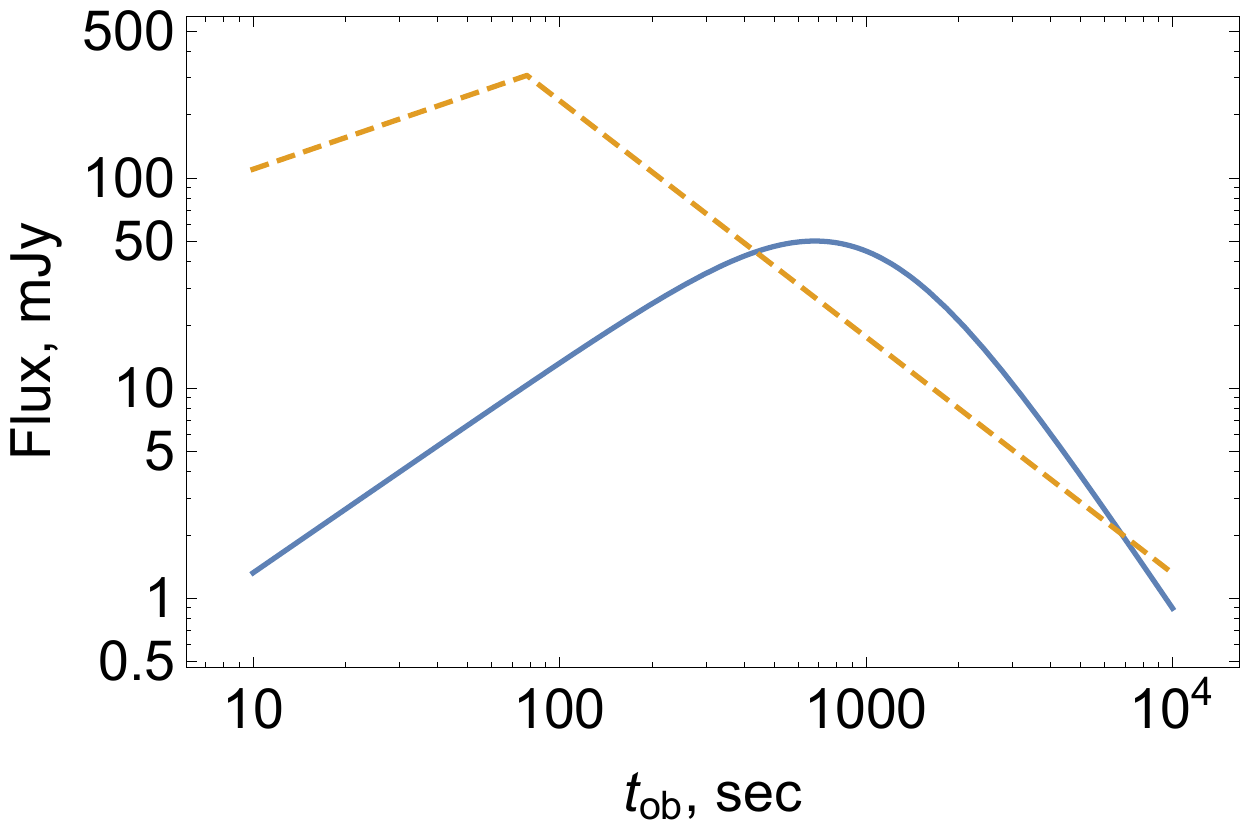}
  \caption{ Evolution of    $X$-ray (left panel)  and optical  (right panel) light  curves for  time-dependent wind luminosity, 
   $L_w= L_{w,0} /(1 + ( t_{ob}/t_0)^2)$ with $t_0 =10^3 $ seconds.  Solid lines are for  RS, dashed lines are for FS. }
 \label{SpectLwt}
\end{figure}

To illustrate the overall behavior,  in Fig. \ref{SpectLwt} we plot the spectra and light curves for assumed wind luminosity $L_w= L_{w,0} /(1 + ( t_{ob}/t_0)^2)$ with $t_0 =10^3 $ seconds. Such power dependence is expected for magnetar-powered wind with spin-down time $t_0$.
For times $t_{ob} \gg t_0$ the time-behavior of the X-ray spectral flux is expected to follow
\be
F_\epsilon \propto \epsilon_m ^{p-1} t_{ob}^{-2} \propto t_{ob}^{ ( 5 p -73)/34} \approx t_{ob}^{-1.78}
\ee
This compares favorably with the post-plateau flux decays.

\begin{figure}[h!]
 \centering
 \includegraphics[width=.99\columnwidth]{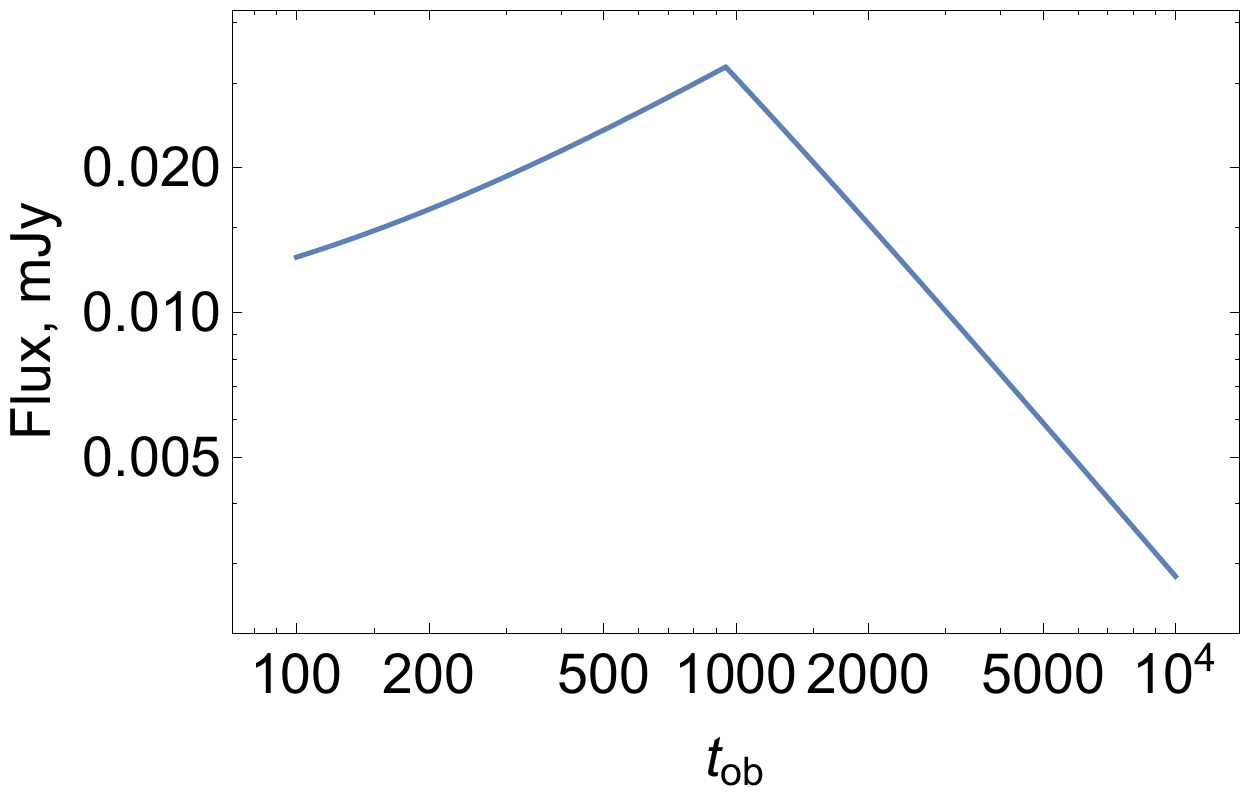}
  \caption{  $X$-ray light curve for varying wind \Lf, $\gamma_w =10^{6} (1+t_{ob}/(100 \rm{sec}))$. The break is associated with the increasing peak energy that passes though detector's energy range. }
 \label{gammawt}
\end{figure}

Variations in the wind \Lf\ $\gamma_w$  also can produce features in the light curves. For example, if $\gamma_w$ increases with time, the peak energy moves to higher values - passing through the detector range it will produce a light curve break (associated with the spectral change from $-p/2$ to $-1/2$, Fig. \ref{gammawt}).

\section{Later afterglows, $t_{ob} \geq E_1/L_w$.}
\label{later}
At times $t_{ob} \geq E_1/L_w$ the wind deposits amount of energy into the FS exceeding the initial explosion
({\it provided}   the source produces  sufficiently high luminosity for sufficiently long time).
 After this time the forward shock is affected by the long lasting engine. 
(This case also applies if the initial explosion was produced by the wind itself.) At this stage the dynamics of the RS changes: now the wind is working against all the swept-up material, so that the CD/RS follow the FS. There is no simple analytical solution  for $f(\chi)$, hence below we just estimate the typical frequencies, not their temporal evolution/light curves.

\begin{figure}[h!]
 \centering
\includegraphics[width=.99\columnwidth]{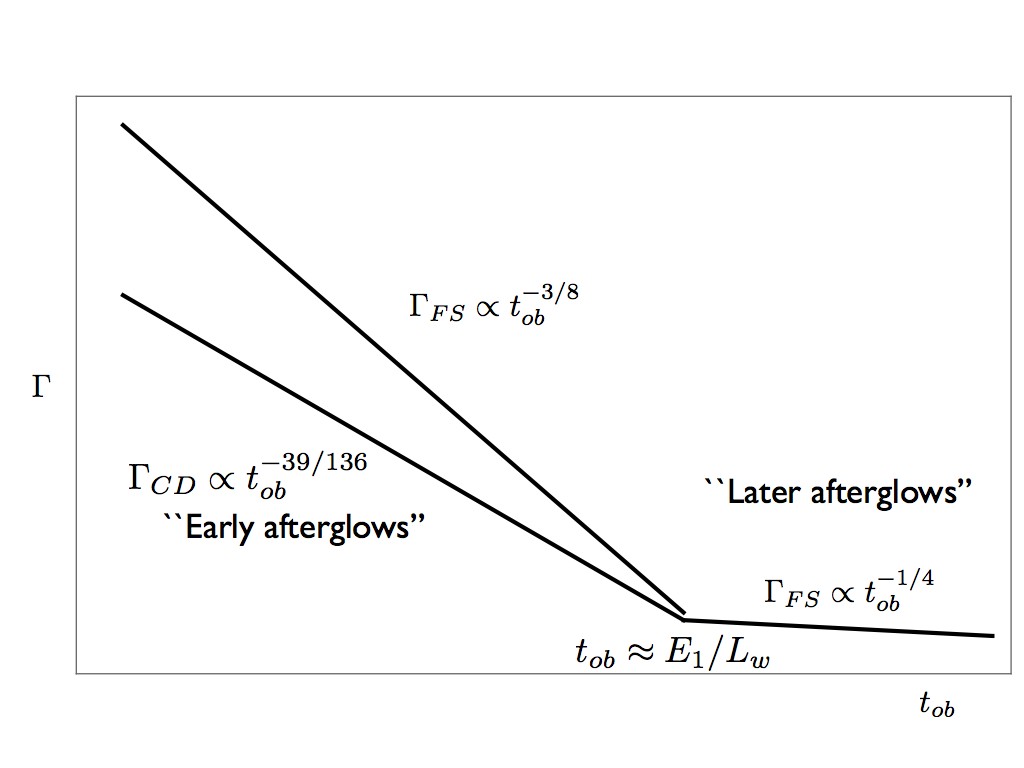}
  \caption{ Temporal evolution of the { \Lf}s of the FS and the CD for constant wind power and constant external density. At time $t_{ob}\approx L_w/E_1$ the second FS merges with the first; at later times the energy deposited by the wind dominates over the initial explosion energy. This changes  the FS from point explosion to constant energy supply.}
 \label{Shock-structure-time}
\end{figure}

The FS flow  follows the self-similar solution of B\&Mc  for blast in constant density with constant energy supply. The FS moves with
\be
\Gamma_{FS} \approx \frac{ \left( {3}/{\pi } \right)^{1/4}}{2} \left (\frac{L_w}{c^5 \rho  t^2}\right)^{1/4}
\approx \left(  \frac{L_w}{c^5 \rho } \right)^{1/8} t_{ob}^{-1/4}
\ee
(Fig. \ref{Shock-structure-time}). 
The \Lf\ of the RS is then
\be
\Gamma_{RS} \approx \Gamma_{FS}/( 2 \sqrt{1+\sigma_w})
\ee
The wind \Lf\ in the frame of the RS is
\be
\gamma_w ' \approx \frac{\gamma_w}{2 \Gamma_{RS}} =  \frac{\gamma_w \sqrt{1+\sigma_w}}{ \Gamma_{FS}} \propto t^{3/2}
\label{gammawlate}
\ee 
Thus the relative  \Lf\ of the wind with respect to the RS (and thus, by assumption, the energy of accelerated particles) {\it increases with time}
(at the catch-up stage it was nearly constant, Eq. (\ref{gammawprime}).
 As a result, the typical synchrotron energy remains constant 
\be
\epsilon_m \approx \frac{\left( {3}/{\pi } \right)^{1/4}}{2} \frac{ e \hbar  }{m_e} \gamma_w^2 \sqrt{\rho} \sqrt{\sigma_w(1+\sigma_w)}
=2\, {\rm keV}\gamma_{w,6}^2 \sqrt{\sigma_w(1+\sigma_w)} n_{ex}^{1/2}
\ee

The cooling time 
   \be
   \tau_c =
  \frac{c^{17/4} m_e^3}{2 e^4} \frac{\sqrt{1+\sigma_w}}{\sigma_w} \frac{t_{ob}^{1/2}}{L_w^{1/4} \gamma_w \rho^{3/4}}
     \ee
     is well shorter than the observer time,
     \be
     \frac{\tau_c}{t_{ob}}\approx 10^{-2} \frac{\sqrt{1+\sigma_w}}{\sigma_w}\frac{1}{L_{w,50}^{1/4} n_{ex}^{3/4} t_{ob,3}^{1/2} \gamma_{w,6}}
     \ee
Since $ \tau_c /t_{ob} \propto t_{ob}^{-1/2}$, the electrons stay in the fast cooling regime (this is because their energy increases with time, $\gamma_e ' \sim \gamma_w '$, (\ref{gammawlate})).
Since $\epsilon_m$ is constant, the spectral power just follows the wind luminosity $L_w$

 \section{Early afterglows from reverse shock emission}

Let us next hypothesize  how emission from the long-lasting RS in fast wind can resolve a number of contradicting GRB observations, contrasting with the FS model, extending the discussion of  \cite{lyutikov_09}. In short, we propose that $X$-rays are coming from the RS, optical is a mixture of RS and FS emission.

{\bf Plateaus.}
In the present model the synchrotron emission is just the wind power, Eq. (\ref{Ls1}). Thus, to produce a plateau all is needed is a $\sim$ temporally constant  wind power.  The total energy produced by the long lasting sources can well be subdominant to the primary explosion. In the present model the plateaus  appear simultaneously in the optical and in the $X$-rays; explaining results of  \cite{2007MNRAS.377.1638D} (both optical and X-ray light curves exhibit a broad plateau).

In contrast,  if associated with FS, this requires that the FS energy  $E_{FS}$ increases with time, $E_{FS} \propto t^{1/2}$. Thus, a central engine should inject $\sim 10$ times more energy at the plateau phase  than was available  during the prompt phase \citep[eg][]{2002ApJ...566..712Z}. Since the total energy in the outflow is constrained by late radio observations, this put unreasonable demands on efficiency of prompt emission.

{\bf Flares}. In the present model the afterglow flares correspond to changing parameters of the wind. These changes affect the emission instantaneously (when a perturbation arrives at the RS). 
 
One of the curious properties of afterglow $X$-ray flares is their short duration, shorter than the expected $\Delta t/t_{ob} \sim 1$ (this is because an instantaneous  emission pulse from radius $r \sim c t$ will be spread  in terms of observational time over $\Delta t \sim t/\Gamma^2 \approx t_{ob}$ due to time-of-flight effects). If a flare is produced by a luminosity pulse of relative strength $L_p = \eta_L L_w$, with $\eta_L \sim 10$, then when it arrives at the RS the \Lf\ of the RS will change instantaneously by $\eta_L^{1/4}$, Eqns (\ref{GammaCD}). 
The corresponding duration will be shorter by $\eta_L^{-1/2}\sim 3$. (Also, at the end of the ``early stage'' the \Lf\ of the RS exceeds the \Lf\ of the FS. 

In contrast, in the framework of FS afterglows, flares correspond to additional injection of energy into the FS. Each flare with the flux increase of the order of unity requires energy injection of the order of total energy already in the outflow. Thus, in case of GRBs with many flares,  the total energy in the FS grows in geometrical progression. 
Also, the models that advocate $X$-flare as late collisions of shells with low \Lf\ \citep[\eg][]{2000ApJ...532..286K} suffer from the efficiency problem. Only the  energy of relative motion  of shells can be converted into radiation. For collision of  low-$\Gamma$ shells this is very inefficient process. \cite{2007MNRAS.375L..46L} concluded that it is problematic to produce flares in the FS. 

{\bf Abrupt endings of the plateau phases}.
One of the most surprising observation of early afterglow is the sudden disappearance of the $X$ emission at the end of the plateau phase \cite[\eg][]{2007ApJ...665..599T}. First, these observations exclude, in our view, the forward shock origin of the plateaus - the forward shock cannot just disappear. Thus, if a central source switches off at coordinate time $t_{off}$, after time $\sim  t_{off} \Gamma_{RS}^2$ the information will reach the RS - it will disappear at the observer time $t_{ob} \sim   t_{off}$. Since the particles are in a fast cooling regime, effective emissivity from the RS will become zero instantaneously. The observed light curve will be modified by the time-of-flight and shock curvature effects. 

In contrast, the reverse shock emissivity can become zero instantaneously.  The observed light curve will decay on a time scale $\sim (r/c)/\Gamma_{RS}^2$. Since $\Gamma_{RS}$ can exceed $\Gamma_{FS}$, the  drop can be sharper than the observer time $t_{ob} \sim (r/c)/\Gamma_{FS}^2$.

 {\bf Fast optical variations.}
There is a number of GRBs that show fast optical variability (\eg\ GRB021004 and most notoriously GRB080916C). This is hard to explain within the standard matter-dominated FS-RS paradigm, since optical electrons should be in a  slow cooling regime (due to mild {\Lf}s of matter-dominated  ``shells'' and small magnetization).

In  the present model the optical emission from the RS  is in the fast cooling regime. Variations of the wind parameters then can produce fast optical variations of the observed flux.

{\bf Naked  GRBs (GRBs without an appreciable afterglow emission).} If the explosion does not produce a long lasting engine, there is no  $X$-ray afterglow.

In the FS paradigm, Naked  GRBs , \eg GRB050421,  are hard to produce in the FS model \citep{KumarPanaitescuNakedGRB}. 

 {\bf Similarity of short and long GRB's afterglows}. In the reverse shock paradigm, both  short and long GRB's afterglows are driven by the highly magnetized/highly relativistic wind, which might  have similar character in both cases: unipolar inductor-driven  light ultra-relativistic pair  wind. 

This similarity  is surprising in the FS  model:  the properties of the forward shock do depend on the external density, while the prompt emission is independent of it. \citep[][give a detailed discussion of evolution of FS spectra; in some cases the dependence on the external density is indeed weak]{2002ApJ...568..820G}

{\bf Missing jet breaks}.
The FS model of afterglows predicts that when the bulk Lorentz factor becomes  $\Gamma \sim 1/\theta_j$ ($\theta_j$ is jet opening angle), an achromatic break should be observed in the light curve \citep{Rhoads}. In some  case there is no jet break \cite[most exciting being 130427A,][with no jet break between few hundred seconds  and 80 Ms]{2016MNRAS.462.1111D}. In other cases breaks are chromatic.

 In the present model, the RS/CD propagate with slowly changing  \Lf\ before the catch-up time (which can be very long for a powerful initial explosion). No jet break is expected then.
Alternatively, for an early jet break,  the ``Later stage", \S \ref{later}, can be applicable: a very long lasting central engine drives a FS-RS configuration, producing emission at the RS  in the fast cooling regime.

{\bf Chromatic jet breaks}. Temporal changes  observed in the light curves - jet breaks - are often interpreted in terms of the morphology of the forward shock \cite{Rhoads}. The prediction is then that such changes   should be achromatic - occurring simultaneously at different frequencies. In fact, often the breaks are chromatic with no break in optical contemporaneous with the $X$-ray break or with different jumps in optical and X-rays, and sometimes breaks are not observed at all, \eg, GRB061007, \citep{Panaitescu2007,Racusin09}.

In the present model the $X$-ray afterglows are coming from the RS, while optical is a mixture of RS and  FS. This allows one to explain achromatic breaks (in cases when optical is dominated by RS and thus is tied to $X$-rays) and chromatic breaks (in cases when optical is dominated by FS and thus is unrelated to $X$-rays). 

{\bf Missing orphan afterglows}. Decelerating FS emits into larger angles at later times. Thus, we should have seen many  orphan afterglows in X-rays and optical - without the prompt emission \cite[eg][]{2002ApJ...570L..61G}. None have been solidly detected, though the rate estimate is not straightforward \citep{2010ApJ...722..235V,2015A&A...578A..71G}. 

In the present model,  afterglow's optical and $X$-emission depends on the details of the RS, \eg on the collimation properties of the fast wind. If the later wind has similar collimation as that of the prompt emission, and since the RS's \Lf\ remains nearly constant in time, fewer orphan afterglows are expected.

{\bf Missing reverse shocks problem}.
The standard model had a clear prediction, of a bright optical flare with a definite decay properties \citep{1996ApJ...473..204S,1997ApJ...476..232M,Sari:1999}. Though a flare closely resembling the predictions was indeed observed \citep[GRB990123,][]{1999MNRAS.306L..39M}, this was an exception.  In the {\it Swift} era,  optical flashes are rare \citep{Gomboc};   even when they are seen, they rarely correspond to RS predictions (often optical emission correlates with X-rays,  too variable, wrong decay laws).

In the present model the early $X$-ray afterglows is the RS emission.

Other miscellaneous points include: (i) The present model requires fast wind, $\gamma_w \geq 10^4$, Eq. (\ref{gammawmin}). It is expected that as the near surrounding of the central engine becomes cleaner with time, the \Lf\ of the wind will increase. If initially $\gamma_w$ is not sufficiently high, the RS emission will be in the slow cooling regime, with smaller  luminosity. As the wind becomes cleaner, \Lf\ $\gamma_w$ increases and it's luminosity increases as well.  
(ii)
We associate smooth optical emission with the FS component. We expect that at the onset of ``Later phase", \S \ref{later}, the dynamics of the FS changes from point explosion to constant wind luminosity. This may be reflected in  flattening of optical light curves at later times. (iii)
The emission from the second FS can also contribute to the observed fluxes. During the early stage its \Lf\ is smaller that  that of the the primary FS, Eq. (\ref{GammaCD}). Hence its emission is generally weaker (but may still be more pronounced at lower energies). On the other hand, the second FS propagates through particles accelerated in the primary shock. The primary shock may work as a injector, making acceleration at the second FS more effective.
(iii) Very early $X$-ray afterglows, with steep temporal decay are probably  the ``high latitude'' tails of prompt emission. (iv) Post-plateau steepening could be due to the declining wind power, \eg\ magnetar spin-down, Fig. \ref{SpectLwt}.
(iv) Infra-red and radio afterglows also could be coming form the RS, but our experience with radio-emitting electrons in pulsar wind nebulae indicate that they form  a different population from the high energy electrons \citep{kennel_coroniti_84b,atoyan_99};  hence, extending simple models of high energy emission to lower frequencies may not be justified. (v) Extended emission in short GRBs \citep{2006ApJ...643..266N,2008MNRAS.385L..10T} may also originate in the reverse shock of long-lasting central engine.

\section{Discussion}

In this paper we argue the observed  temporal and spectral properties of early $X$-ray afterglows are generally consistent with the synchrotron emission coming from particles accelerated behind a 
RS propagating  in a long-lasting ultra-high relativistic wind. Since the late wind is expected to be fast and highly magnetized, the RS emission is very efficient, converting into radiation a  large fraction of the wind power.
 This high efficiency of conversion of wind power into radiation does not place tough constraints on the wind luminosity and total power. In the simplest model, the plateaus are very flat in time, Eq. (\ref{Fnu}). 

The main points of the work are:
\begin{itemize}
\item Since the circumburst cavity is cleared by the preceding forward blast and since the reverse shock is powered by the central engine, the \Lf\ of the RS behaves very differently from the FS, typically decreasing with a slower rate than  that of the FS.
\item In the fast cooling regime the RS synchrotron  emission is, approximately, the wind power. Thus, in order to power afterglows a mildly luminous central source is needed.   
   \item The  $X$-ray spectral flux   is nearly constant, Eq.  (\ref{Fnu}). This provides  a natural explanation for the early $X$-ray plateaus. 
   \item In the optical, contribution from FS and RS are approximately comparable. The FS is expected to produce smooth light curves (in the slow cooling regime), while the optical emission of  RS  (in the fast cooling regime, Eq. (\ref{ecc})) may show fast variability correlated with $X$-rays. The optical emission from the RS can also produce plateaus.
 \end{itemize}
 
Importantly,  RS luminosity depends on the instantaneous wind power at the location of the RS, and thus can be highly variable \citep[\eg due to development of instabilities in the jet][]{2011MNRAS.411L..16L}.  In contrast to the FS emission, which depends on the total integrated energy absorbed by the media; thus in order to produce, \eg flares in the FS model one must change the {\it total energy}  of the FS 
   shock. Also, fast cooling regime ensures  that at any moment the observed emission corresponds to the instantaneous shock power (modulo time-of-flight effects).
   Thus, flares, as well as   sudden dimming of the afterglows  \cite[akin to the one discussed by][]{2007ApJ...665..599T} are more naturally explained in the RS scenario. For example, if the wind from the central engine terminates, the RS disappears the moment that the back part of the flow reached it. (It is nearly impossible to image how a FS emission can terminate nearly instantaneously).  We expect that variations in $L_w$ and  $\sigma_w$ will be able to  account for a variety of early afterglow behaviors.
 
 There is  a number of uncertain parameters in the model (\eg wind luminosity, \Lf\ and  magnetization), as well as theoretical limitations. 
 The model requires sufficiently high  wind luminosity and \Lf; magnetization could be mild, but high \Lf\ requires high magnetization. The main theoretical limitation is in treating the second shock in the thin  shell approximation. We expect that more realistic calculation (\eg\ numerical simulations)  would produce slightly different time evolution for $\Gamma_{CD}$. But the present model  only weakly depends on the particular time-dependence of $\Gamma_{CD}$: higher $\Gamma_{CD}$ would  decrease the post-shock \Lf\ of the wind particles, but, on the other hand, would  increase the boost to the observer frame. This may change the particular time scalings, but as long as the RS emission remains in the fast cooling regime,  our key  estimates will remain true. To illustrate this point (that the  RS  emission properties are roughly independent of the assumed approximations for strong RS, in Appendix \ref{Comp} we calculate the RS dynamics using Kompaneets approximation - balancing wind and the post-shock pressures - and in Appendix \ref{delayed}  we calculate the RS with a delayed start; though the particularities are different, the main conclusions stands).

In conclusion, the emission from the reverse shock of long-lived engine that produces highly relativistic wind is (i) highly efficient (due to high \Lf\  of the wind and   high \Bf\  supplied by the central source); (ii) stable for constant wind parameters (its \Lf\ is nearly constant as a function of time); (iii) can react quickly to the changes of the wind properties (and thus can explain rapid variability). All these properties point to the RS emission as the origin of early $X$-ray afterglows. RS also contributes to optical - this explains correlated X-optical features often seen in afterglows. FS emission occurs in the optical range, and, at later times, in radio.

We would like to thank Maxim Barkov,  Edo Berger, Rodolfo Barniol Duran, Hendrik van Eerten, Gabriele Ghisellini, Dimitrios Giannios, Robert Preece and Eleonora Troja  for discussions and comments on the manuscript. 

This work had been supported by   NSF  grant AST-1306672 and DoE grant DE-SC0016369.

\bibliographystyle{apj}
  \bibliography{/Users/maxim/Home/Research/BibTex}

\begin{thebibliography}{62}
\expandafter\ifx\csname natexlab\endcsname\relax\def\natexlab#1{#1}\fi

\bibitem[{{Atoyan}(1999)}]{atoyan_99}
{Atoyan}, A.~M. 1999, \aap, 346, L49

\bibitem[{{Barkov} \& {Komissarov}(2010)}]{2010MNRAS.401.1644B}
{Barkov}, M.~V., \& {Komissarov}, S.~S. 2010, \mnras, 401, 1644

\bibitem[{{Barkov} \& {Pozanenko}(2011)}]{2011MNRAS.417.2161B}
{Barkov}, M.~V., \& {Pozanenko}, A.~S. 2011, \mnras, 417, 2161

\bibitem[{{Bisnovatyi-Kogan} \& {Silich}(1995)}]{Bisnovatyi-KoganSilich95}
{Bisnovatyi-Kogan}, G.~S., \& {Silich}, S.~A. 1995, Reviews of Modern Physics,
  67, 661

\bibitem[{{Blandford} \& {McKee}(1976)}]{blandford_76}
{Blandford}, R.~D., \& {McKee}, C.~F. 1976, Physics of Fluids, 19, 1130

\bibitem[{{de Pasquale} {et~al.}(2007){de Pasquale}, {Oates}, {Page},
  {Burrows}, {Blustin}, {Zane}, {Mason}, {Roming}, {Palmer}, {Gehrels}, \&
  {Zhang}}]{2007MNRAS.377.1638D}
{de Pasquale}, M., {Oates}, S.~R., {Page}, M.~J., {et~al.} 2007, \mnras, 377,
  1638

\bibitem[{{De Pasquale} {et~al.}(2016){De Pasquale}, {Page}, {Kann}, {Oates},
  {Schulze}, {Zhang}, {Cano}, {Gendre}, {Malesani}, {Rossi}, {Troja}, {Piro},
  {Bo{\"e}r}, {Stratta}, \& {Gehrels}}]{2016MNRAS.462.1111D}
{De Pasquale}, M., {Page}, M.~J., {Kann}, D.~A., {et~al.} 2016, \mnras, 462,
  1111

\bibitem[{{Fong} {et~al.}(2014){Fong}, {Berger}, {Metzger}, {Margutti},
  {Chornock}, {Migliori}, {Foley}, {Zauderer}, {Lunnan}, {Laskar}, {Desch},
  {Meech}, {Sonnett}, {Dickey}, {Hedlund}, \& {Harding}}]{2014ApJ...780..118F}
{Fong}, W., {Berger}, E., {Metzger}, B.~D., {et~al.} 2014, \apj, 780, 118

\bibitem[{{Gehrels} \& {Razzaque}(2013)}]{2013FrPhy...8..661G}
{Gehrels}, N., \& {Razzaque}, S. 2013, Frontiers of Physics, 8, 661

\bibitem[{{Genet} {et~al.}(2007){Genet}, {Daigne}, \&
  {Mochkovitch}}]{2007MNRAS.381..732G}
{Genet}, F., {Daigne}, F., \& {Mochkovitch}, R. 2007, \mnras, 381, 732

\bibitem[{{Ghirlanda} {et~al.}(2015){Ghirlanda}, {Salvaterra}, {Campana},
  {Vergani}, {Japelj}, {Bernardini}, {Burlon}, {D'Avanzo}, {Melandri},
  {Gomboc}, {Nappo}, {Paladini}, {Pescalli}, {Salafia}, \&
  {Tagliaferri}}]{2015A&A...578A..71G}
{Ghirlanda}, G., {Salvaterra}, R., {Campana}, S., {et~al.} 2015, \aap, 578, A71

\bibitem[{{Ghisellini} {et~al.}(2007){Ghisellini}, {Ghirlanda}, {Nava}, \&
  {Firmani}}]{2007ApJ...658L..75G}
{Ghisellini}, G., {Ghirlanda}, G., {Nava}, L., \& {Firmani}, C. 2007, \apjl,
  658, L75

\bibitem[{{Gomboc}(2009)}]{Gomboc}
{Gomboc}, A.~{\etal}. 2009, in American Institute of Physics Conference Series,
  Vol. 1133, American Institute of Physics Conference Series, ed. {C.~Meegan,
  C.~Kouveliotou, \& N.~Gehrels}, 145--150

\bibitem[{{Granot} {et~al.}(2002){Granot}, {Panaitescu}, {Kumar}, \&
  {Woosley}}]{2002ApJ...570L..61G}
{Granot}, J., {Panaitescu}, A., {Kumar}, P., \& {Woosley}, S.~E. 2002, \apjl,
  570, L61

\bibitem[{{Granot} \& {Sari}(2002)}]{2002ApJ...568..820G}
{Granot}, J., \& {Sari}, R. 2002, \apj, 568, 820

\bibitem[{{Icke}(1988)}]{Icke}
{Icke}, V. 1988, \aap, 202, 177

\bibitem[{{Kann} {et~al.}(2010){Kann}, {Klose}, {Zhang}, {Malesani}, {Nakar},
  {Pozanenko}, {Wilson}, {Butler}, {Jakobsson}, {Schulze}, {Andreev},
  {Antonelli}, {Bikmaev}, {Biryukov}, {B{\"o}ttcher}, {Burenin}, {Castro
  Cer{\'o}n}, {Castro-Tirado}, {Chincarini}, {Cobb}, {Covino}, {D'Avanzo},
  {D'Elia}, {Della Valle}, {de Ugarte Postigo}, {Efimov}, {Ferrero}, {Fugazza},
  {Fynbo}, {G{\aa}lfalk}, {Grundahl}, {Gorosabel}, {Gupta}, {Guziy}, {Hafizov},
  {Hjorth}, {Holhjem}, {Ibrahimov}, {Im}, {Israel}, {Je{\'l}inek}, {Jensen},
  {Karimov}, {Khamitov}, {Kizilo{\v g}lu}, {Klunko}, {Kub{\'a}nek}, {Kutyrev},
  {Laursen}, {Levan}, {Mannucci}, {Martin}, {Mescheryakov}, {Mirabal},
  {Norris}, {Ovaldsen}, {Paraficz}, {Pavlenko}, {Piranomonte}, {Rossi},
  {Rumyantsev}, {Salinas}, {Sergeev}, {Sharapov}, {Sollerman}, {Stecklum},
  {Stella}, {Tagliaferri}, {Tanvir}, {Telting}, {Testa}, {Updike}, {Volnova},
  {Watson}, {Wiersema}, \& {Xu}}]{2010ApJ...720.1513K}
{Kann}, D.~A., {Klose}, S., {Zhang}, B., {et~al.} 2010, \apj, 720, 1513

\bibitem[{{Katz} {et~al.}(1998){Katz}, {Piran}, \&
  {Sari}}]{1998PhRvL..80.1580K}
{Katz}, J.~I., {Piran}, T., \& {Sari}, R. 1998, Physical Review Letters, 80,
  1580

\bibitem[{{Kennel} \& {Coroniti}(1984{\natexlab{a}})}]{kennelCoroniti84}
{Kennel}, C.~F., \& {Coroniti}, F.~V. 1984{\natexlab{a}}, \apj, 283, 694

\bibitem[{{Kennel} \& {Coroniti}(1984{\natexlab{b}})}]{kennel_84}
---. 1984{\natexlab{b}}, \apj, 283, 710

\bibitem[{{Kennel} \& {Coroniti}(1984{\natexlab{c}})}]{kennel_coroniti_84b}
---. 1984{\natexlab{c}}, \apj, 283, 710

\bibitem[{{Komissarov} \& {Barkov}(2009)}]{2009MNRAS.397.1153K}
{Komissarov}, S.~S., \& {Barkov}, M.~V. 2009, \mnras, 397, 1153

\bibitem[{{Kompaneets}(1960)}]{Komp}
{Kompaneets}, A.~S. 1960, Soviet Physics Doklady, 5, 46

\bibitem[{{Kumar} \& {Panaitescu}(2000)}]{KumarPanaitescuNakedGRB}
{Kumar}, P., \& {Panaitescu}, A. 2000, \apjl, 541, L51

\bibitem[{{Kumar} \& {Piran}(2000)}]{2000ApJ...532..286K}
{Kumar}, P., \& {Piran}, T. 2000, \apj, 532, 286

\bibitem[{{Kumar} \& {Zhang}(2015)}]{2015PhR...561....1K}
{Kumar}, P., \& {Zhang}, B. 2015, \physrep, 561, 1

\bibitem[{{Laskar} {et~al.}(2013){Laskar}, {Berger}, {Zauderer}, {Margutti},
  {Soderberg}, {Chakraborti}, {Lunnan}, {Chornock}, {Chandra}, \&
  {Ray}}]{2013ApJ...776..119L}
{Laskar}, T., {Berger}, E., {Zauderer}, B.~A., {et~al.} 2013, \apj, 776, 119

\bibitem[{{Lazzati} {et~al.}(2011){Lazzati}, {Blackwell}, {Morsony}, \&
  {Begelman}}]{2011MNRAS.411L..16L}
{Lazzati}, D., {Blackwell}, C.~H., {Morsony}, B.~J., \& {Begelman}, M.~C. 2011,
  \mnras, 411, L16

\bibitem[{{Lazzati} \& {Perna}(2007)}]{2007MNRAS.375L..46L}
{Lazzati}, D., \& {Perna}, R. 2007, \mnras, 375, L46

\bibitem[{{Lien} {et~al.}(2016){Lien}, {Sakamoto}, {Barthelmy}, {Baumgartner},
  {Cannizzo}, {Chen}, {Collins}, {Cummings}, {Gehrels}, {Krimm}, {Markwardt},
  {Palmer}, {Stamatikos}, {Troja}, \& {Ukwatta}}]{2016ApJ...829....7L}
{Lien}, A., {Sakamoto}, T., {Barthelmy}, S.~D., {et~al.} 2016, \apj, 829, 7

\bibitem[{{Lyutikov}(2006)}]{2006NJPh....8..119L}
{Lyutikov}, M. 2006, New Journal of Physics, 8, 119

\bibitem[{{Lyutikov}(2009)}]{lyutikov_09}
---. 2009, ArXiv e-prints 0911.0349

\bibitem[{{Lyutikov}(2011)}]{2011MNRAS.411.2054L}
---. 2011, \mnras, 411, 2054

\bibitem[{{Lyutikov}(2013)}]{2013ApJ...768...63L}
---. 2013, \apj, 768, 63

\bibitem[{{Lyutikov} {et~al.}(2016){Lyutikov}, {Komissarov}, \&
  {Porth}}]{2016MNRAS.456..286L}
{Lyutikov}, M., {Komissarov}, S.~S., \& {Porth}, O. 2016, \mnras, 456, 286

\bibitem[{{Lyutikov} \& {McKinney}(2011)}]{2011PhRvD..84h4019L}
{Lyutikov}, M., \& {McKinney}, J.~C. 2011, \prd, 84, 084019

\bibitem[{{M{\'e}sz{\'a}ros}(2006)}]{MeszarosReview}
{M{\'e}sz{\'a}ros}, P. 2006, Reports on Progress in Physics, 69, 2259

\bibitem[{{Meszaros} \& {Rees}(1997)}]{1997ApJ...476..232M}
{Meszaros}, P., \& {Rees}, M.~J. 1997, \apj, 476, 232

\bibitem[{{M{\'e}sz{\'a}ros} \& {Rees}(1999)}]{1999MNRAS.306L..39M}
{M{\'e}sz{\'a}ros}, P., \& {Rees}, M.~J. 1999, MNRAS, 306, L39

\bibitem[{{Metzger} {et~al.}(2011){Metzger}, {Giannios}, {Thompson},
  {Bucciantini}, \& {Quataert}}]{2011MNRAS.413.2031M}
{Metzger}, B.~D., {Giannios}, D., {Thompson}, T.~A., {Bucciantini}, N., \&
  {Quataert}, E. 2011, \mnras, 413, 2031

\bibitem[{{Norris} \& {Bonnell}(2006)}]{2006ApJ...643..266N}
{Norris}, J.~P., \& {Bonnell}, J.~T. 2006, \apj, 643, 266

\bibitem[{{Nousek} {et~al.}(2006){Nousek}, {Kouveliotou}, {Grupe}, {Page},
  {Granot}, {Ramirez-Ruiz}, {Patel}, {Burrows}, {Mangano}, {Barthelmy},
  {Beardmore}, {Campana}, {Capalbi}, {Chincarini}, {Cusumano}, {Falcone},
  {Gehrels}, {Giommi}, {Goad}, {Godet}, {Hurkett}, {Kennea}, {Moretti},
  {O'Brien}, {Osborne}, {Romano}, {Tagliaferri}, \&
  {Wells}}]{2006ApJ...642..389N}
{Nousek}, J.~A., {Kouveliotou}, C., {Grupe}, D., {et~al.} 2006, \apj, 642, 389

\bibitem[{{Paczynski}(1986)}]{Paczynski86}
{Paczynski}, B. 1986, \apjl, 308, L43

\bibitem[{{Panaitescu}(2007)}]{Panaitescu2007}
{Panaitescu}, A. 2007, MNRAS, 380, 374

\bibitem[{{Piran}(1999)}]{PiranReview}
{Piran}, T. 1999, \physrep, 314, 575

\bibitem[{{Piran}(2004)}]{piran_04}
---. 2004, Reviews of Modern Physics, 76, 1143

\bibitem[{{Racusin} {et~al.}(2009){Racusin}, {Liang}, {Burrows}, {Falcone},
  {Sakamoto}, {Zhang}, {Zhang}, {Evans}, \& {Osborne}}]{Racusin09}
{Racusin}, J.~L., {Liang}, E.~W., {Burrows}, D.~N., {et~al.} 2009, \apj, 698,
  43

\bibitem[{{Rees} \& {Meszaros}(1992)}]{MeszarosRees92}
{Rees}, M.~J., \& {Meszaros}, P. 1992, MNRAS, 258, 41P

\bibitem[{{Rhoads}(1997)}]{Rhoads}
{Rhoads}, J.~E. 1997, \apjl, 487, L1+

\bibitem[{{Sari} {et~al.}(1996){Sari}, {Narayan}, \&
  {Piran}}]{1996ApJ...473..204S}
{Sari}, R., {Narayan}, R., \& {Piran}, T. 1996, \apj, 473, 204

\bibitem[{{Sari} \& {Piran}(1995)}]{Sari95}
{Sari}, R., \& {Piran}, T. 1995, \apjl, 455, L143+

\bibitem[{{Sari} \& {Piran}(1999)}]{Sari:1999}
---. 1999, ApJ, 517, L109

\bibitem[{{Sari} {et~al.}(1998){Sari}, {Piran}, \&
  {Narayan}}]{1998ApJ...497L..17S}
{Sari}, R., {Piran}, T., \& {Narayan}, R. 1998, \apjl, 497, L17

\bibitem[{{Silva} {et~al.}(2003){Silva}, {Fonseca}, {Tonge}, {Dawson}, {Mori},
  \& {Medvedev}}]{2003ApJ...596L.121S}
{Silva}, L.~O., {Fonseca}, R.~A., {Tonge}, J.~W., {et~al.} 2003, \apjl, 596,
  L121

\bibitem[{{Spitkovsky}(2008)}]{2008ApJ...673L..39S}
{Spitkovsky}, A. 2008, \apjl, 673, L39

\bibitem[{{Troja} {et~al.}(2008){Troja}, {King}, {O'Brien}, {Lyons}, \&
  {Cusumano}}]{2008MNRAS.385L..10T}
{Troja}, E., {King}, A.~R., {O'Brien}, P.~T., {Lyons}, N., \& {Cusumano}, G.
  2008, \mnras, 385, L10

\bibitem[{{Troja} {et~al.}(2007){Troja}, {Cusumano}, {O'Brien}, {Zhang},
  {Sbarufatti}, {Mangano}, {Willingale}, {Chincarini}, {Osborne}, {Marshall},
  {Burrows}, {Campana}, {Gehrels}, {Guidorzi}, {Krimm}, {La Parola}, {Liang},
  {Mineo}, {Moretti}, {Page}, {Romano}, {Tagliaferri}, {Zhang}, {Page}, \&
  {Schady}}]{2007ApJ...665..599T}
{Troja}, E., {Cusumano}, G., {O'Brien}, P.~T., {et~al.} 2007, \apj, 665, 599

\bibitem[{{Uhm} \& {Beloborodov}(2007)}]{Uhm}
{Uhm}, Z.~L., \& {Beloborodov}, A.~M. 2007, \apjl, 665, L93

\bibitem[{{Usov}(1992)}]{Usov92}
{Usov}, V.~V. 1992, \nat, 357, 472

\bibitem[{{van Eerten} {et~al.}(2010){van Eerten}, {Zhang}, \&
  {MacFadyen}}]{2010ApJ...722..235V}
{van Eerten}, H., {Zhang}, W., \& {MacFadyen}, A. 2010, \apj, 722, 235

\bibitem[{{Yuan} \& {Blandford}(2015)}]{2015MNRAS.454.2754Y}
{Yuan}, Y., \& {Blandford}, R.~D. 2015, MNRAS, 454, 2754

\bibitem[{{Zhang} \& {M{\'e}sz{\'a}ros}(2002)}]{2002ApJ...566..712Z}
{Zhang}, B., \& {M{\'e}sz{\'a}ros}, P. 2002, \apj, 566, 712

\end{thebibliography}

\appendix
\section{RS dynamics in the  Kompaneets approximation}
\label{Comp}

The  Kompaneets approximation \citep{Komp} balances post-shock and ram pressure  (for non-relativistic application in astrophysics see \citep[\eg][]{Icke}; relativistic  case was discussed by
 \cite{ 2011MNRAS.411.2054L}).  Since the post-first shock plasma is relativistically hot and moving away from the center of explosion, this ram pressure depends on the local enthalpy 
and the relative \Lf\ between the local plasma flow and the second shock/the CD. We can estimate both those values using the self-similar solution for a relativistic point explosion (B\&Mc).

The pressure balance on the CD  between the wind and the immediate pre-shock ram pressure is
\ba &&
\frac{L_w} {4\pi r^2 c \Gamma_{CD}^2}= w_{CD} \left( \frac{\Gamma_{CD}}{2\gamma _{CD}}\right)^2 c^2=
\frac{8}{3} \Gamma_{CD}^2 \rho c^2 \chi_{CD}^{7/12}
\label{force}
\\ &&
 w_{CD}= 4 \times \frac{2}{3} \rho \Gamma_{FS}^2  \chi_{CD}^{-17/12}
 \nn &&
 \gamma_{CD} \approx   \Gamma_{FS}/(2 \chi)
 \label{pressureCDA}
\ea
where $ w_{CD}$ is the enthalpy and $\gamma_{CD}$ is the post-first shock flow right in front of the second shock/the CD.
Here on the left  is the ram pressure created by the freely expanding wind; it is smaller by  $\Gamma_{CD}^2$ in the frame of the CD. On the right is the ram pressure of the second FS, which is  approximately the pressure in the post second shock fluid. The ram pressure is the enthalpy (since the primary shocked fluid has $p \gg  \rho$)  times the  (\Lf)$^2$ of the motion of the CD with respect to the upstream fluid. This \Lf\  is, approximately, the  Lorentz factor of the CD, $\Gamma_{CD}$, divided by the Lorentz factor of the fluid in front of it, $\gamma_{CD}$. (One of the tests of this scaling is the Blandford-McKee scaling for the FS with energy supply; in that case the  enthalpy equals the  outside density, the Lorentz factor of the fluid in front of it is 1, which gives  $ \Gamma_{CD} \propto  t^{-1/2}$.) Importantly,  the Lorentz factor of the primary shock (and of the pre-second FS flow at the location of the CD, $\gamma_{CD}$) cancels  out: pre-second shock enthalpy $\propto\gamma_{CD}^ 2 \propto \Gamma_{FS}^2$, but the ram pressure that  the second shock creates is $\propto\gamma_{CD}^ {-2} \propto   \Gamma_{FS}^{-2}$. (We remind, the small $\gamma_{CD}$ refers to the \Lf\ of the fluid in front of the CD, while the capital   $\Gamma_{CD}$ 
refers to the \Lf\ of the CD.)

Solving (\ref{force}) for $\Gamma_{CD}$, we find
\be
\Gamma_{CD}\approx  \frac{L_w^{1/4}} { c^{3/4} r_{CD}^{1/2} \rho^{1/4} \chi_{CD}^{7/48}}
\label{gammaCDa}
\ee
This gives the \Lf\ of the CD between the doubly-shock external medium and the constant power wind with luminosity $L_w$; the CD is located at the self-similar coordinate $\chi_{CD}$ of the primary shock.

Combining (\ref{rCD}) and (\ref{gammaCDa})
we find the self-similar dynamics of the second shock:   $m=3/17$. The \Lf\ and the location of the CD are then given by
\ba && 
\Gamma_{CD} = 0.25  \left( \frac{L_w^{12}}{E_1^7 c^{25} \rho ^5} \right)^{1/34} t^{-3/34}
\nn &&
\chi_{CD}\approx 10.9 \left(\frac{L_w t}{ E_1 \Gamma_{FS}^2}\right)^{-12/17} 
\label{GammaCDa}
\ea
In terms of the observer time,
\be
\Gamma_{CD} = 0.4  \frac{L_w^{3/10}}{E_1^{7/40} c^{5/8} {\rho } ^{1/8}{t_{ob}}^{3/40}}
\ee
The corresponding peak energy is 
\be
\epsilon_m \approx 1.7 \times 10^4 \, {\rm eV} 
\frac{E_{1,54}^{7/10} \sqrt{{n_{ex}}} \gamma _{w,6}^2}{L_{w,50}^{7/10}
   t_{{ob},3}^{7/10}}
   \label{epsilons11}
   \ee
   The peak energy  (\ref{epsilons11}) has, qualitatively - and given the uncertainties of the parameters - the same order-of-magnitude value as the estimate (\ref{epsilons1}).
The cooling energy evaluates to 
   \be
\epsilon_c \approx  10^{-3} \, {\rm eV} 
\frac{L_{w,50}^{9/10} t_{{ob},3}^{2/5}}{E_{1,54}^{7/5} {n_{ex}}} \ll \epsilon_m
\ee
Thus, though the dynamics of the CD in the Kompaneets approximation  is somewhat different, the key conclusions  remain valid Ð the $X$-ray emission is in the fast cooling regime. 
    
   \section{RS dynamics with delayed start}
\label{delayed}

Another scaling for the dynamics  of the second shock is possible if the onset of the  wind  is delayed with respect to the initial explosion.
Suppose that the second explosion/secondary wind occurs at  time $t_d$ after the initial one and the second shock/CD  is moving with the  \Lf\ $\Gamma_{CD} ^2 \propto (t-t_d)^{-m} \gg \Gamma_1 ^2$.  Then, the location of the second shock at time $t$ is 
\be
R_{CD} = (t-t_d) \left( 1-\frac{1}{2 \Gamma _{CD}^2 (m+1)} \right)
\ee
The corresponding self-similar coordinate of the second shock in terms of the primary shock self-similar parameter $\chi$  is 
\be
\chi_{CD} =(1 +8 \Gamma_1^2 ) (1-\frac{R}{t}) 
\approx
\left( \frac{8 t_d}{t} + \frac{4}{(m+1) \Gamma_{CD}^2} \right) \Gamma _1^2
\ee
For $t_d \geq t/(2 (m+1) \Gamma_{CD}^2)$ the location of the CD in the self-similar coordinate associated with the first shock is 
\be
\chi_{CD} \approx
\frac{8 \Gamma _1^2  {t_d}}{t}\propto t^{-4}
\label{chi2}
\ee
In this case
\ba &&
\Gamma_{CD} =0.52 \frac{E_1^{5/48} {t_d}^{5/48} {L_w}^{1/4}}{c^{85/48} \rho ^{17/48} t^{11/12}}= 0.46 \frac{E_1^{5/136} {t_d}^{5/136} L_w^{3/34}}{c^{5/8} {\rho }^{1/8}
   t_{{ob}}^{11/34}} 
   \nn &&
   \chi_{CD}=2.6 \frac{E_1^{24/17}}{c^{60/17} \rho ^{12/17} t^{48/17} L_w^{12/17}}=8. 2 \frac{E_1^{12/5}}{L_w^{12/5} t_{{ob}}^{12/5}}
   \ea
   The peak energy,
   \be
   \epsilon_m \approx 2.6 \times 10^3\, {\rm eV} \frac{\sqrt{{n_{ex}}} L_{w,50}^{5/34} t_{{ob},3}^{5/17} \gamma
   _{w,6}^2}{E_{1,54}^{5/34} {t_d}^{5/34}}
   \ee
   is much larger than the cooling frequency
     \be
   \epsilon_c \approx 0.2 \, {\rm eV}\frac{E_{1,54}^{5/17} {t_d}^{5/24}}{{n_{ex}} L_{w,50}^{27/34}
   t_{{ob},3}^{613/408}}
      \ee
   
     We conclude that various approximations for the second shock dynamics do not affect our main conclusion (that the RS  in the  fast wind is a very efficient in converting the wind energy into the high energy radiation).

 \section{Wind environment}
 \label{wind}
 
 
 For the wind environment $\rho =\rho_0 (r/r_0)^{-2}$, the initial blast wave
 propagates according to
 \be
 \Gamma_{FS} = \frac{3}{2 \sqrt{2 \pi} c^{3/2}} \sqrt{\frac{E_1}{\rho_0 r_0^2 t}}
 \label{GammaFSw}
 \ee 
and  creates a profile (B\&Mc) $f \propto \chi^{-3/2}$, $g \propto 1/\chi$. 
Accumulated enthalpy, 
\be
w_2 = \frac{16 \pi}{3} \frac{r_0^2 c^3 \rho_0 t}{\sqrt{\chi_{CD}}},
\ee
gives the \Lf\ of the CD
\be
\Gamma_{CD} =0.41 \frac{L_w^{1/4} \chi_{CD}^{1/8}}{c^{3/4} \sqrt{r_0} \rho_0^{1/4}}
\ee
Together with (\ref{rCD}) and (\ref{GammaFSw}) we find
\ba &&
\Gamma_{CD}= 0.49 \frac{{E_1}^{1/10} {L_w}^{1/5}}{c^{9/10} \rho _0^{3/10} r_0^{3/5} {t}} 
=0.55
\frac{{E_1}^{1/12} {L_w}^{1/6}}{c^{3/4} {\rho _0} ^{1/4}\sqrt{r_0}
{t_{{ob}}^{1/12}}}
   \nn &&
   \chi_{CD} = 1.46 \frac{E_1^{4/5}}{c^{6/5} \rho _0^{2/5} r_0^{4/5} t^{4/5} L_w^{2/5}}=
   3.7 \left(\frac{E_1}{L_w t_{{ob}}}\right){}^{2/3}
   \nn &&
   t= 0.31\frac{{E_1} ^{1/6} {L_w} ^{1/3}t_{{ob}}^{5/6}}{c^{3/2} \sqrt{\rho _0} r_0}
   \ea
   The peak and the cooling energies evolve according  to
   \ba && 
   \epsilon_m = 5.2 \frac{ \sqrt{c} e \rho _0 r_0^2 \sqrt{\sigma_w } \sqrt{1+\sigma_w} \hbar  \gamma
   _w^2}{{E_1}^{1/3} m_e {L_w}^{1/6} t_{{ob}}^{2/3}}
   \nn &&
    \epsilon_c= 10^{-2}
    \frac{ E_1^{2/3} c^{15/2} m_e^5 (1+\sigma_w)^{3/2} \hbar 
 {t_{ob}^{1/3}}}{e^7 \rho _0^2 r_0^4 \sigma_w ^{3/2}{L_w}^{1/6}}
   \ea

At later times, when the CD catches with the FS, we find
\ba
&&
\Gamma_{CD}\approx \Gamma_{FS} \approx \frac{L_w^{1/4}}{ c^{3/4} \rho_0^{1/4} r_0^{1/2}}
\nn &&
\epsilon_m = \frac{ e \hbar}{m_e c} \sqrt{\sigma_w(1+\sigma_w)} 
\frac{\sqrt{\rho _0} r_0 \gamma _w^2}{t} =\frac{\sqrt{c} e \rho _0 r_0^2 \sqrt{\sigma_w  (\sigma_w +1)} \hbar  \gamma _w^2}{m_e
   \sqrt{L_w} t_{{ob}}}
   \nn &&
   \epsilon_c=\frac{c^{15/2} m_e^5 \hbar  \sqrt{L_w} t_{{ob}}}{e^7 \rho _0^2 r_0^4
   }\left(\frac{\sigma_w }{\sigma_w +1}\right)^{-3/2}
   \ea

 \end{document}